\documentclass{jfm}
\usepackage{newtxtext}
\usepackage{newtxmath}

\usepackage[T1]{fontenc}
\usepackage[english]{babel}%
\usepackage{csquotes}%
\usepackage[activate={true,nocompatibility},tracking=true]{microtype}%
\usepackage{hyphenat}%
\usepackage{siunitx}
\usepackage{placeins}
\usepackage[begintext=``, endtext='']{quoting}
\SetBlockEnvironment{quoting}
\SetBlockThreshold{1}
\usepackage{hyperref}

\let\proof\relax

\usepackage{amsthm}
\usepackage{amsmath}
\usepackage{amsfonts}
\usepackage{amssymb}
\usepackage{amscd}
\usepackage{mathtools}
\usepackage[mathscr]{eucal}
\usepackage{calligra}
\DeclareMathAlphabet{\mathcalligra}{T1}{calligra}{m}{n}
\DeclareFontShape{T1}{calligra}{m}{n}{<->s*[1.1]callig15}{}
\usepackage{bm}
\usepackage[nice]{nicefrac}
\usepackage{dsfont}

\usepackage{tikz}
\usepackage{pgf}
\usepackage{pgfplots}
\usepackage{pgfplotstable}
\usepackage{shellesc}
\pgfplotsset{compat=newest}
\usetikzlibrary{spy}
\usetikzlibrary{shapes}
\usetikzlibrary{backgrounds}
\usetikzlibrary{shadows}
\usetikzlibrary{matrix}
\usetikzlibrary{external}
\usetikzlibrary{positioning}
\usetikzlibrary{plotmarks}
\usepgfplotslibrary{fillbetween}
\usepackage{natbib}
\usepackage{comment}
\usepackage{graphicx}
\usepackage[colorinlistoftodos,prependcaption,textsize=small]{todonotes}
\usepackage[capitalize]{cleveref}

\crefname{equation}{}{}
\crefname{figure}{Figure}{Figures}
\usepackage{csvsimple}
\usepackage{booktabs}
\usepackage{environ}
\usepackage{adjustbox}
\usepackage{relsize}
\usepackage{afterpage}
\usepackage{booktabs}
\usepackage{arydshln}

\newtheorem{remark}{Remark}[section]

\newcommand{\RomanNumeralCaps}[1]
\linenumbers

\let\min\relax \DeclareMathOperator*\min{\vphantom{p}min}
\let\max\relax \DeclareMathOperator*\max{\vphantom{p}max}
\let\subset\relax \DeclareMathOperator{\subset}{\subseteq}
 
\let\tilde\widetilde
\let\hat\widehat

\DeclareMathOperator{\grad}{grad}

\DeclareMathOperator{\curl}{curl}
\let\div\relax %
\DeclareMathOperator{\div}{div}

\newcommand{\sspace}{\hspace{0.25pt}}

\newcommand{\todoBKinline}[1]{\tikzexternaldisable\todo[color=red!10,inline]{\footnotesize{\bf Brendan:} #1}\tikzexternalenable}

\newcommand{\R}{\mathbb{R}} %
\newcommand{\Z}{\mathbb{Z}} %
\newcommand{\dd}{\,\mathrm{d}} %
\newcommand{\bdry}{\partial} %
\newcommand{\spann}{\mathrm{span}} %
\newcommand{\bzero}{\boldsymbol{0}} %

\newcommand{\bmeta}{{\bm{\eta}}}
\newcommand{\btheta}{{\bm{\theta}}}

\newcommand{\bxi}{\bm{\xi}}

\newcommand{\bphi}{\bm{\phi}}

\newcommand{\bpsi}{\bm{\psi}}

\newcommand{\bomega}{\bm{\omega}}
\newcommand{\bTheta}{\bm{\Theta}}

\newcommand{\bma}{\bm{a}}
\newcommand{\bmb}{\bm{b}}

\newcommand{\bme}{\bm{e}}

\newcommand{\bmk}{\bm{k}}

\newcommand{\bmn}{\bm{n}}

\newcommand{\bmr}{\bm{r}}

\newcommand{\bmv}{\bm{v}}
\newcommand{\bmw}{\bm{w}}
\newcommand{\bmx}{\bm{x}}
\newcommand{\bmy}{\bm{y}}

\newcommand{\bmU}{\bm{U}}

\newcommand{\bmW}{\bm{W}}

\newcommand{\bmY}{\bm{Y}}
\newcommand{\bmZ}{\bm{Z}}

\newcommand{\sfd}{\mathsf{d}}
\newcommand{\sfe}{\mathsf{e}}
\newcommand{\sff}{\mathsf{f}}

\newcommand{\sfp}{\mathsf{p}}

\newcommand{\bsfb}{\bm{\mathsf{b}}}

\newcommand{\bsfd}{\bm{\mathsf{d}}}

\newcommand{\bsff}{\bm{\mathsf{f}}}

\newcommand{\bsfp}{\bm{\mathsf{p}}}

\newcommand{\sfI}{\mathsf{I}}

\newcommand{\sfQ}{\mathsf{Q}}

\newcommand{\bsfA}{\mathsfbi{A}}
\newcommand{\bsfB}{\mathsfbi{B}}

\newcommand{\bsfD}{\mathsfbi{D}}

\newcommand{\bsfG}{\mathsfbi{G}}
\newcommand{\bsfH}{\mathsfbi{H}}
\newcommand{\bsfI}{\mathsfbi{I}}

\newcommand{\bsfL}{\mathsfbi{L}}
\newcommand{\bsfM}{\mathsfbi{M}}

\newcommand{\bsfQ}{\mathsfbi{Q}}

\newcommand{\bsfT}{\mathsfbi{T}}

\newcommand{\mcD}{\mathcal{D}}

\newcommand{\mcF}{\mathcal{F}}

\newcommand{\mcJ}{\mathcal{J}}

\newcommand{\mcL}{\mathcal{L}}
\newcommand{\mcM}{\mathcal{M}}
\newcommand{\mcN}{\mathcal{N}}

\newcommand{\bbE}{\mathbb{E}}

\newcommand{\bbP}{\mathbb{P}}

\newcommand{\bbZ}{\mathbb{Z}}

\newcommand{\bfj}{\mathbf{j}}

\newcommand{\bfu}{\mathbf{u}}

\newcommand{\bfw}{\mathbf{w}}

\newcommand{\bfA}{\mathbf{A}}

\newcommand{\bfU}{\mathbf{U}}

\newcommand{\bfW}{\mathbf{W}}

\newcommand{\bfZ}{\mathbf{Z}}

\newcolumntype{?}{!{\vrule width 1.2pt}}

\makeatletter
\newsavebox{\measure@tikzpicture}
\NewEnviron{scaletikzpicturetowidth}[1]{%
	\tikzifexternalizingnext{%
		\def\tikz@width{#1}%
		\begin{lrbox}{\measure@tikzpicture}%
			\tikzset{external/export next=false,external/optimize=false}%
			\BODY
		\end{lrbox}%
		\pgfmathparse{#1/\wd\measure@tikzpicture}%
		\BODY
	}{%
		\BODY
	}%
}
\makeatother

\definecolor{color0}{rgb}{0.7843, 0.7843, 0.7843}
\definecolor{color1}{rgb}{0, 0.4470, 0.7410}
\definecolor{color2}{rgb}{0.8500, 0.3250, 0.0980}
\definecolor{color3}{rgb}{0.9290, 0.6940, 0.1250}
\definecolor{color4}{rgb}{0.7060, 0.3840, 0.7650}
\definecolor{color5}{rgb}{0.4660, 0.6740, 0.1880}
\definecolor{color6}{rgb}{0.3010, 0.7450, 0.9330}
\definecolor{color7}{rgb}{0.6350, 0.0780, 0.1840}
\definecolor{color8}{rgb}{0.0, 0.4078, 0.3412}

\pgfplotsset{
  log x ticks with fixed point/.style={
      xticklabel={
        \pgfkeys{/pgf/fpu=true}
        \pgfmathparse{exp(\tick)}%
        \pgfmathprintnumber[fixed relative, precision=3]{\pgfmathresult}
        \pgfkeys{/pgf/fpu=false}
      }
  },
  log y ticks with fixed point/.style={
      yticklabel={
        \pgfkeys{/pgf/fpu=true}
        \pgfmathparse{exp(\tick)}%
        \pgfmathprintnumber[fixed relative, precision=3]{\pgfmathresult}
        \pgfkeys{/pgf/fpu=false}
      }
  }
}

\tikzset{
  ashadow/.style={opacity=.25, shadow xshift=0.07, shadow yshift=-0.07},
}

\definecolor{CustomGreen}{RGB}{65,169,50}

\renewcommand{\vec}[1]{\bm{#1}}
\newcommand{\ten}[1]{\bm{#1}}
\newcommand{\drv}[2]{\frac{\partial #1}{\partial #2}}
\renewcommand{\grad}{\nabla}
\renewcommand{\div}{\operatorname{\nabla\hspace{1.25pt}\bcdot}}
\renewcommand{\curl}{\operatorname{\nabla\times}}
\newcommand{\abs}[1]{\left| #1 \right|}
\newcommand{\inv}{^{\scalebox{0.7}[1.0]{\( - \)}1}}%

\newcommand{\avg}[1]{\langle #1 \rangle}
\newcommand{\ii}{\operatorname{i}}
\newcommand{\ee}{\operatorname{e}}

\newcommand{\Fourier}{\mathcal{F}}

\newcommand{\pot}{\vec{\psi}} %
\newcommand{\Id}{I}%

\newcommand{\VonKarman}{von K{\'a}rm{\'a}n}

\newcommand{\Matern}{\mathcal{M}}
\newcommand{\BesselK}{K}

\pgfplotsset{
    legend image with text/.style={
        legend image code/.code={%
            \node[anchor=center] at (0.3cm,0cm) {#1};
        }
    },
}

\makeatletter
\tikzset{
    scale plot marks/.is choice,
    scale plot marks/false/.code={
        \def\pgfuseplotmark##1{\pgftransformresetnontranslations\csname pgf@plot@mark@##1\endcsname}
    },
    scale plot marks/true/.style={},
    scale plot marks/.default=true
}
\makeatother \graphicspath{{./Figures/}}

\hypersetup{colorlinks=true,citecolor=black,filecolor=black,linkcolor=black,urlcolor=black}
\hypersetup{pdfstartview=FitB,pdfpagemode=UseNone}

\title{A fractional PDE model for turbulent velocity fields near solid walls}

\newtheorem{lemma}{Lemma}

\author{Brendan~Keith
  \corresp{\email{keith@ma.tum.de}},
  Ustim~Khristenko
 \and Barbara~Wohlmuth}

\begin{document}

\maketitle

\begin{abstract}
	This paper presents a class of turbulence models written in terms of fractional partial differential equations (FPDEs) with stochastic loads.
	Every solution of these FPDE models is an incompressible velocity field and the distribution of solutions is Gaussian.
	Interaction of the turbulence with solid walls is incorporated through the enforcement of various boundary conditions.
	The various boundary conditions deliver extensive flexibility in the near-wall statistics that can be modelled.
	Reproduction of both fully-developed shear-free and uniform shear boundary layer turbulence are highlighted as two simple physical applications; the first of which is also directly validated with experimental data. 
	The rendering of inhomogeneous synthetic turbulence inlet boundary conditions is an additional application, motivated by contemporary numerical wind tunnel simulations.
	Calibration of model parameters and efficient numerical methods are also conferred upon.
\end{abstract} 

\section{Introduction} %
\label{sec:introduction}

Solid walls and other boundaries have a variety of well-known effects on turbulent flows.
This paper is concerned with forming a statistical model which incorporates many of these effects and can be used to efficiently generate independent identically distributed synthetic turbulent velocity fields.
These random velocity fields can then be employed in uncertain quantification (UQ) for computational fluid dynamics (CFD), wherein random velocity fields are typically used as simulation inputs, or, for example, the generation of the synthetic turbulent boundary conditions, as we demonstrate within.

The statistical model we propose is a boundary value problem with a stochastic right-hand side and a (non-local) fractional differential operator with two fractional exponents.
The exponents determine the shape of the energy spectrum in the energy-containing range and the inertial subrange, while the regularity of the right-hand side specifies the shape of the dissipative range.
Finally, the choice of boundary conditions and other model parameters shape the spatial dependence of the energy spectra near the solid boundary.

If the stochastic load appearing on the right-hand side is Gaussian, then the turbulence model will deliver a Gaussian distributed random velocity field (GRVF) with zero mean and an implicitly defined covariance tensor.
Gaussian random fields (GRFs) are essentially ubiquitous in contemporary UQ and many convenient features of them are well-known; see, e.g., \citet{Liu2019} and references therein.
In particular, fractional differential operators and other types of non-local operators are important tools which may be used to represent a wide variety of random field models.
Notable recent advances in fluid mechanics involving such operators include \citet{chen2006speculative,song2018universal,mehta2019discovering,egolf2019nonlinear,di2020two}.
Each of these works mainly focus on extensions of RANS and LES models.
Here, we focus directly on modelling and generating turbulent velocity field fluctuations.

The Fourier transform can be used to characterize homogeneous turbulence and it may, of course, also be used directly to generate synthetic velocity fields; see, e.g., \citet{mann1998wind}.
Various models for such spectral tensors have been investigated to describe homogeneous velocity fields for various conditions; cf. \citet{hinze1959Turbulence,Maxey1982,kristensen1989spectral,mann1994spatial}.
The seminal work of Hunt et al. \citep{hunt1973theory,Hunt1978free,Hunt1984turbulence} describes a relatively simple procedure to amend these homogeneous models, making them inhomogeneous and applicable to the inviscid source layer around a large impenetrable body.
The class of models presented here can be seen as an extension of Hunt's original ideas.
The most obvious departure between the two approaches, however, is that ours involves characterizing a vector potential which is, in turn, post-processed to deliver the synthetic turbulence.
Meanwhile, Hunt's approach, briefly summarized in the next section, involves post-processing the original homogeneous velocity field by removing a conservative and solenoidal vector field term.

In~\cref{sec:preliminaries,sec:general_models}, we derive a general fractional partial differential equation (FPDE) model for the stochastic vector potential $\bpsi$.
On simply connected domains, the expression
\begin{equation}
\label{eq:VectorPotential}
	\bfu
	=
	\nabla \times \bpsi
	,
\end{equation}
then immediately defines the corresponding (incompressible) turbulent fluctuations $\bfu$.
In~\cref{sec:preliminaries}, the well-known von K\'arm\'an energy spectrum \citep{von1948progress} is used as a motivating example.
This preliminary model is then embellished throughout~\cref{sec:general_models}; for example, via a detailed analysis of first-order shearing effects and through the assignment of boundary conditions.
Various applications of the turbulence models are discussed in~\cref{sec:applications}, including its use in generating synthetic turbulence inlet boundary conditions.
In~\cref{sec:calibration_and_rational_approximation}, numerical methods and model calibration are briefly surveyed and, finally, the complete findings are summarized in~\cref{sec:conclusions}.

\section{Motivation for a vector potential model} %
\label{sec:relationship_to_previous_work}

Before entering the main body of this paper, we briefly review Hunt's classical approach to the construction of inhomogeneous turbulence near solid walls \citep{Hunt1984turbulence,nieuwstadt2016turbulence}.
We denote $z>0$ as the distance from the wall, $\nu$ as the kinematic viscosity, $L_\infty$ as the integral length scale, and $\bfu^{(H)}$ as homogeneous turbulence, distributed everywhere in space in the same way that the turbulent velocity field $\bfu$ is far away from the wall.
Moreover, here and throughout, $\langle \,\cdot\, \rangle$ denotes ensemble averaging.

Let $\Omega = \{ (x,y,z) \in \R^3 \colon z>0\}$.
In the inviscid source layer above a infinite solid wall $\bdry\Omega = \{ (x,y,z) \in \R^3 \colon z=0\}$, we have the following idealized boundary conditions on the turbulent velocity field $\bfu$:
\begin{equation*}
	\bfu\bcdot\bmn = 0
	~\text{ as } \frac{z}{L_\infty} \to 0
	,
	\qquad
	\bfu \to \bfu^{(H)}
	~\text{ as } \frac{z}{L_\infty} \to \infty
	\,.
\end{equation*}
Here, $\bmn = \bme_3$ represents the unit normal to $\bdry\Omega$.
In, e.g., a shear-free turbulent layer, both the energy dissipation rate $\epsilon$ and the mean velocity are approximately constant with the height above the surface.
Nevertheless, the turbulent fluctuations $\bfu$ are affected by the boundary.

We now consider the following decomposition:
\begin{equation}
\label{eq:TurbulenceDecomposition}
	\bfu
	=
	\bfu^{(H)} + \bfu^{(S)}
	.
\end{equation}
Here, $\bfu^{(H)}$ denotes the background turbulence in the absence of the boundary, and $\bfu^{(S)}$ denotes the residual fluctuations produced in the inviscid source layer.
Note that such a decomposition introduces an analogous decomposition of the vorticity; namely,
\begin{equation}
\label{eq:VortictyDecomposition}
	\bomega
	=
	\nabla \times\bfu^{(H)} + \nabla \times\bfu^{(S)}
	=
	\bomega^{(H)} + \bomega^{(S)}
	.
\end{equation}

One can show that in the limit $\Rey \to \infty$ \cite[p.~42]{townsend1980structure},
\begin{equation*}
	\epsilon
	=
	\nu \langle\sspace |\bomega|^2 \rangle
	.
\end{equation*}
Therefore, under the idealized assumption $\epsilon = \text{const.}$, the residual vorticity term $\bomega^{(S)}$ may be taken as equal to zero.
It is then natural to assume
\begin{equation}
\label{eq:PotentialFunction}
	\bfu^{(S)} = -\nabla \phi
	,
\end{equation}
for some potential function $\nabla^2 \phi = 0$ in $\Omega$ and $\nabla \phi \bcdot \bmn = \bfu^{(H)} \bcdot \bmn$ on $\bdry\Omega$.
Alternatively, one may consider the more general vector potential representation of $\bfu^{(S)}$:
\begin{equation}
\label{eq:VectorPotentialA}
	\bfu^{(S)}
	=
	-\nabla \times \bfA
	,
\end{equation}
where $-\nabla^2\bfA = \bomega^{(S)}$ and $\div \bfA = 0$ in $\Omega$ and $(\nabla\times\bfA)\bcdot\bmn = \bfu^{(H)} \bcdot \bmn$ and $\bfA\bcdot\bmn = 0$ on $\bdry\Omega$; cf. \cite[Theorem~3.5]{Girault1986}.
Clearly, when $\bomega^{(S)} = 0$, it holds that $\nabla \phi = \nabla \times \bfA$.

A shortcoming of expression~\cref{eq:PotentialFunction} compared to~\cref{eq:VectorPotentialA} is that~\cref{eq:PotentialFunction} is only viable when $\bomega^{(S)} = 0$, however,~\cref{eq:VectorPotentialA} is viable for any $\bomega^{(S)}$.
Likewise, $\bfu^{(H)}$ may always be expressed as the curl of a vector potential, but, generally, cannot be expressed as the gradient of any scalar potential.

From now on, we completely dispense with the idealized assumption $\bomega^{(S)} = 0$ and cease to scrutinize the potential benefits of decompositions~\cref{eq:TurbulenceDecomposition,eq:VortictyDecomposition}.
In short, we simply choose to write $\bfu = \nabla \times \bpsi$, as in~\cref{eq:VectorPotential}, for some vector potential $\bpsi$, which does not necessarily have to be incompressible.
This expression is an essential ingredient in deriving the fractional PDE-based model below.

\section{Preliminaries} %
\label{sec:preliminaries}

In this section, we introduce the main notation of the paper and connect a class free space random fields to solutions of certain FPDEs with a stochastic right-hand side.
In order to ease the presentation in the following section, which pushes this relationship much further, we demonstrate the FPDE connection with an explicit example coming from the Von K\'arm\'an energy spectrum function.

\subsection{Definitions} %
\label{sub:definitions}

We wish to model turbulent velocity fields $\bfU(\bmx) = \langle\bfU(\bmx)\rangle + \bfu(\bmx) \in \R^3$.
Here, $\langle\bfU\rangle = (\langle U_1\rangle,\langle U_2\rangle,\langle U_3\rangle)$ is the mean velocity field and $\bfu = (u_1,u_2,u_3)$ (sometimes also written $(u,v,w)$) are the zero-mean turbulent fluctuations.
All of the models we choose to consider for $\bfu$ are \emph{Gaussian}.
That is, they are determined entirely from the two-point correlation tensor
\begin{equation*}
	R_{ij}(\bmr,\bmx,t)
	=
	\langle u_i(\bmx,t)u_j(\bmx+\bmr,t)\rangle
	.
\label{eq:CovarianceTensor}
\end{equation*}
When $R(\bmr,\bmx,t) = R(\bmr,t)$ depends only on the separation vector $\bmr$, the model is said to be spatially \emph{homogeneous}.
Alternatively, when $R(\bmr,\bmx,t) = R(\bmr,\bmx)$ is independent of the time variable $t$, the model is said to be temporally \emph{stationary}.

Frequently, it is convenient to consider the Fourier transform of the velocity field $\bfu$.
In such cases, we express the field in terms of a generalized Fourier--Stieltjes integral,
\begin{equation}
	\bfu(\bmx)
	=
	\int_{\R^3}
	\ee^{\ii\bmk\bcdot\bmx}
	\dd \bmZ(\bmk)
	\,,
\label{eq:FourierStieltjes}
\end{equation}
where $\bmZ(\bmk)$ is a three-component measure on $\R^3$.
The validity of this expression follows from the Wiener--Khinchin theorem \citep{lord2014introduction}.
Likewise, in the homogeneous setting, we may consider the Fourier transform of the covariance tensor, otherwise known as the \emph{velocity-spectrum tensor},
\begin{equation*}
\label{eq:StationarySpectralTensor}
\Phi_{ij}(\bmk,t)
=
\frac{1}{(2\upi)^3}\int_{\R^3} \ee^{-\ii\bmk\bcdot\bmr}R_{ij}(\bmr,t)\dd \bmr
.
\end{equation*}

Consider three-dimensional additive white Gaussian noise~\citep{hida2013white,kuo2018white} in the physical and frequency domains, denoted $\bxi(\bmx)$ and $\hat{\bxi}(\bmk)$, respectively, such that
\begin{equation*}\label{eq:ksi}
	\bxi(\bmx)
	=
	\int_{\R^3} \ee^{\ii\bmk\bcdot\bmx} \hat{\bxi}(\bmk) \dd \bmk
	=
	\int_{\R^3} \ee^{\ii\bmk\bcdot\bmx} \dd \bfW(\bmk),
\end{equation*}
where $\bmW(\bmk)$ is three-dimensional Brownian motion.
We assume $\dd \bmZ(\bmk) = \bsfG(\bmk) \dd \bmW(\bmk) = \bsfG(\bmk) \hat{\bxi}(\bmk) \dd \bmk$, where $\bsfG(\bmk)^\ast \bsfG(\bmk) = \Phi(\bmk)$.

This section and the next are devoted to deriving fractional PDE models for homogeneous turbulence.
The approach we follow involves a commonly used definition of fractional differential operators facilitated by the spectral theorem~\citep{reed2012methods}.
Note that, for an abstract closed normal operator~$A\colon \mcD(A)\subset H\to H$ on a complex Hilbert space~$H$, $AA^\ast = A^\ast A$, there exists a finite measure space $(Y, \mu)$, together with a complex-valued measurable function $\lambda(y)$, defined on $Y$, and a unitary map $U : H\to L_2(Y, \mu)$, such that
\begin{equation*}
	UA\phi= \lambda U\phi
	\quad
	\text{for all }
	\phi\in H
	.
\end{equation*}
In this case, one may define the $\alpha$-fractional power of $A$ as follows:
\begin{equation}
	A^\alpha= U^\ast \lambda^\alpha U
	.
\label{eq:AbstractSpectralRepresentation}
\end{equation}
For an operator $A\colon\mcD(A)\subset L^2(\Omega)\to L^2(\Omega)$ with a discrete spectrum, we may simply write
\begin{equation}
\label{eq:SpectramTheorem}
	A^{\alpha}\phi
	=
	\sum_{j=1}^\infty
	\lambda_j^{\alpha} (\phi,e_j)_\Omega\, e_j
	.
\end{equation}
Here, $e_j$ and $\lambda_j$ denote the corresponding eigenmodes and eigenvalues of $A$ and $(\phi,\chi)_\Omega = \int_\Omega \phi\bcdot\chi \dd x$ denotes the $L^2$-inner product on the domain $\Omega\subset \R^3$.

For example, consider the vector Laplacian operator $A = -\Delta$ on $\Omega = \R^d$.
Letting $k = |\bmk|$ denote the magnitude of the wavenumber vector $\bmk = (k_1,k_2,k_3)$ in Fourier space and $\mcF$ and $\mcF^{-1}$ denote the Fourier and inverse Fourier transforms, respectively, we have
\begin{equation*}
	(-\Delta)^{\alpha}\bphi(\bmx)
	=
	\frac{1}{(2\upi)^d}
	\int_{\R^d}
	k^{2\alpha}\sspace (\bphi,\ee^{-\ii\bmk\bcdot \bmx})_{\R^d}\,\ee^{\ii\bmk\bcdot \bmx} \dd \bmk
	=
	\mcF^{-1}\{k^{2\alpha}\mcF\{\bphi\}(\bmk)\}(\bmx)
	.
\end{equation*}
Evidently, in this setting, $\mcF$ is the analogue of the unitary operator $U$ present in the abstract expression~\cref{eq:AbstractSpectralRepresentation}.
On the other hand, when $\Omega = (0,1)^d$ is a periodic domain, it is well known that $A = -\Delta$ has a discrete spectrum.
Here, recall that
\begin{equation*}
	(-\Delta)^{\alpha}\bphi(\bmx)
	=
	\frac{1}{(2\upi)^d}
	\sum_{\bfj\in\bbZ^d}
	k_\bfj^{2\alpha} (\bphi,\ee^{-\ii\bmk_\bfj\bcdot \bmx})_{\R^d}\,\ee^{\ii\bmk_\bfj\bcdot \bmx}
	.
\end{equation*}
For further details on the spectral representation of closed operators, we refer the interested reader to \citet{de1975spectral,weidmann2012linear,kowalski2009spectral}.

\subsection{The von K\'{a}rm\'{a}n model} %
\label{sub:energy_spectrum_function}

Let us begin with a standard form of the spectral tensor used in isotropic stationary and homogeneous turbulence models, namely,
\begin{equation}
	\Phi_{ij}(\bmk)
	=
	(4\upi)^{-1}k^{-2} E(k)
	P_{ij}(\bmk)
	\,.
\label{eq:IsotropicSpectralTensor}
\end{equation}
Here, $E(k)$ is called the \emph{energy spectrum function} and $P_{ij}(\bmk) = \delta_{ij} - \frac{k_ik_j}{k^2}$ is commonly referred to as the \emph{projection tensor}.
One common empirical model for $E(k)$, suggested by \citet{von1948progress}, is given by the expression
\begin{equation}
	E(k)
	=
	c_0^2\sspace \varepsilon^{2/3}k^{-5/3}
	\bigg(
	\frac{k L}{(1 + (k L)^2)^{1/2}}
	\bigg)^{17/3}
	.
\label{eq:VKEnergySpectrum}
\end{equation}
Here, $\varepsilon$ is the viscous dissipation of the turbulent kinetic energy, $L$ is a length scale parameter, and $c_0^2\approx 1.7$ is an empirical constant.

Recall that the Fourier transform of the scalar Laplacian is simply $-k^2$.
Likewise, consider the Fourier transform $\bsfQ(\bmk)$ of the $\mathrm{curl}$ operator, $\int_{\R^3} \nabla \times \bmv(\bmr) \ee^{-\ii\bmk\bcdot\bmr}\dd\bmr = \bsfQ(\bmk)\hat{\bmv}(\bmk)$, where $\hat{\bmv}(\bmk) = \int_{\R^3} \bmv(\bmr) \ee^{-\ii\bmk\bcdot\bmr}\dd\bmr$.
Observe that
\begin{equation*}
	\bsfQ(\bmk)
	=
	\ii
	\begin{bmatrix}
		0 & -k_3 & k_2\\
		k_3 & 0 & -k_1\\
		-k_2 & k_1 & 0
	\end{bmatrix}
\end{equation*}
and, moreover, $P(\bmk) = k^{-2}\bsfQ(\bmk)^\ast \bsfQ(\bmk)$.
Motivated by the decomposition $\Phi(\bmk) = \bsfG(\bmk)^\ast \bsfG(\bmk)$, we choose to simply write
	$
	\bsfG(\bmk)
	=
	\frac{1}{\sqrt{4\upi}}
	k^{-2} E^{1/2}(k)
	\bsfQ(\bmk)
	.
	$
Next, recalling $\dd \bmZ(\bmk) = \bsfG(\bmk) \dd \bmW(\bmk)$, it immediately follows that
\begin{equation*}
	\dd \bmZ(\bmk)
	=
	\bsfQ(\bmk) \Big(\frac{1}{\sqrt{4\upi}k^2} E^{1/2}(k) \dd\bmW(\bmk)\Big)
	.
\end{equation*}
Integrating both sides with respect to $\bmk$, we arrive at the expression $\bfu = \nabla \times \bpsi$,
with a vector potential defined
\begin{equation}
\label{eq:VectorPotentialFourier}
	\bpsi(\bmx)
	=
	\frac{1}{\sqrt{4\upi}}
	\int_{\R^3} k^{-2} E^{1/2}(k) \ee^{\ii\bmk\bcdot\bmx} \dd \bmW(\bmk)
 	\,.
\end{equation}

We now proceed to relate the vector potential $\bpsi(\bmx)$ to the solution of a fractional PDE.
Writing $\bpsi(\bmx) = \int_{\R^3} \ee^{\ii\bmk\bcdot\bmx} \dd \bmY(\bmk)$, similar to~\cref{eq:FourierStieltjes}, and rearranging the factors in~\cref{eq:VectorPotentialFourier}, leads to
\begin{equation*}
	(1 + (k L)^2)^{17/12} \dd \bmY(\bmk)
	=
	c_0\varepsilon^{1/3}L^{17/6}\dd \bmW(\bmk)
	.
\end{equation*}
Then, upon integrating both sides with respect to $\bmk$, we arrive at the fractional PDE
\begin{equation}
	(\Id-L^2\Delta)^{17/12} \bpsi = c_0 \varepsilon^{1/3} L^{17/6} \bxi
	.
\label{eq:VKFractionalPDE}
\end{equation}
This and all future differential equations are only properly understood in the sense of distributions, yet we continue to use the ``strong form'' for readability.

Let $\Id$ denote the identity operator, $A = \Id-L^2\Delta$, $\mu = c_0 \varepsilon^{1/3}$, and $\alpha = 17/12$.
With these symbols in hand, the derivation above can be summarized as follows:
\begin{equation*}
\label{eq:velocity_from_vpotential}
	\bfu = \curl\bpsi,
	\qquad
	\text{where}\quad
	A^{\alpha} \bpsi = \mu L^{2\alpha} \bxi
	.
\end{equation*}
In the next section, we extend the simple FPDE model above in order to describe inhomogeneous turbulence on bounded domains.
This is achieved by both generalizing the definition of the length scale $L$ and the fractional operator $A^{\alpha}$ as well as introducing a physical notion of boundary conditions.
Before we begin, we remark on the two former aspects.

\begin{figure}
\centering
	\includegraphics[clip=true, trim = 21cm 1cm 22cm 6cm, width=0.28\textwidth]{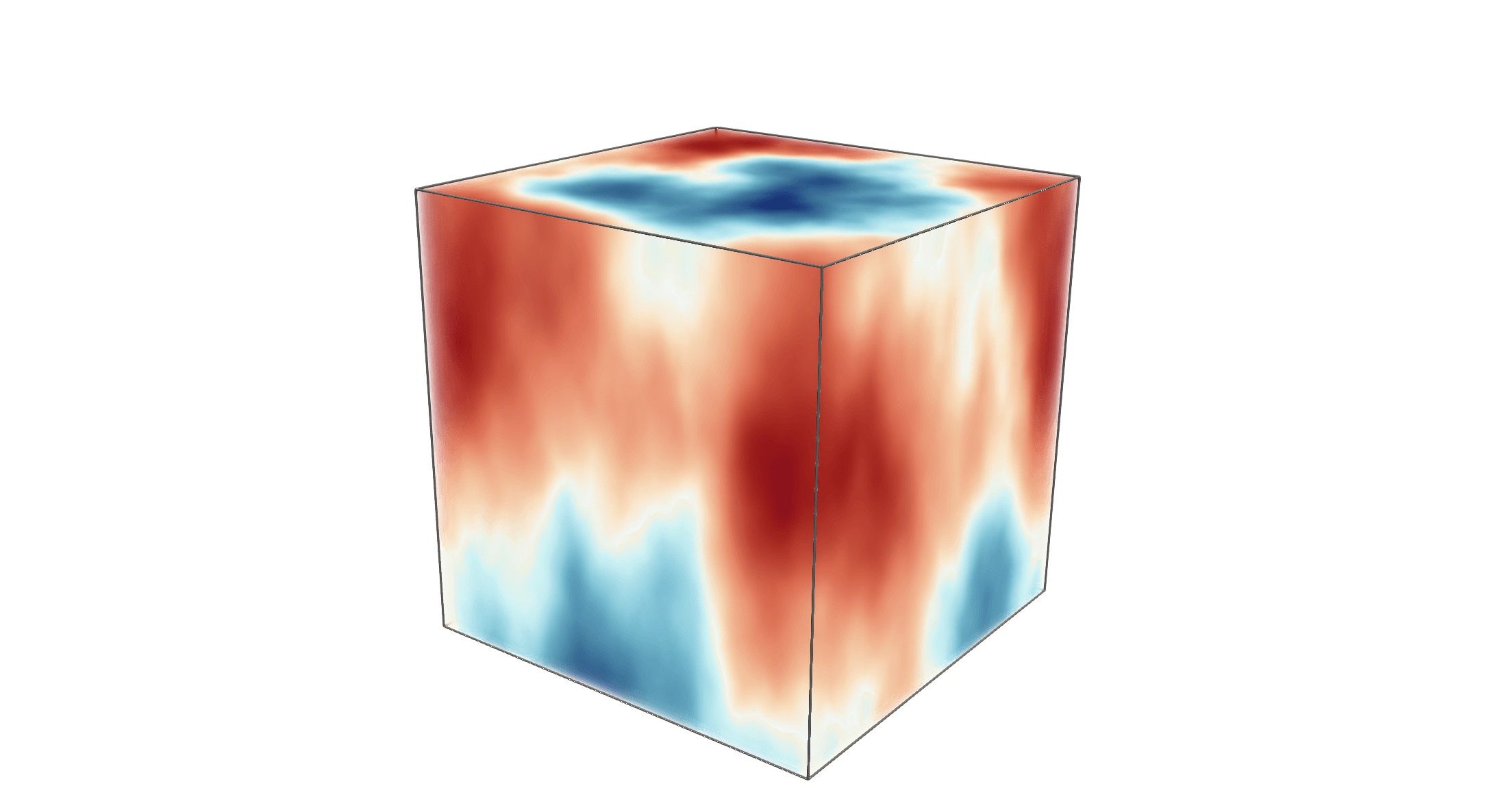}
	~
	\includegraphics[clip=true, trim = 21cm 1cm 22cm 6cm, width=0.28\textwidth]{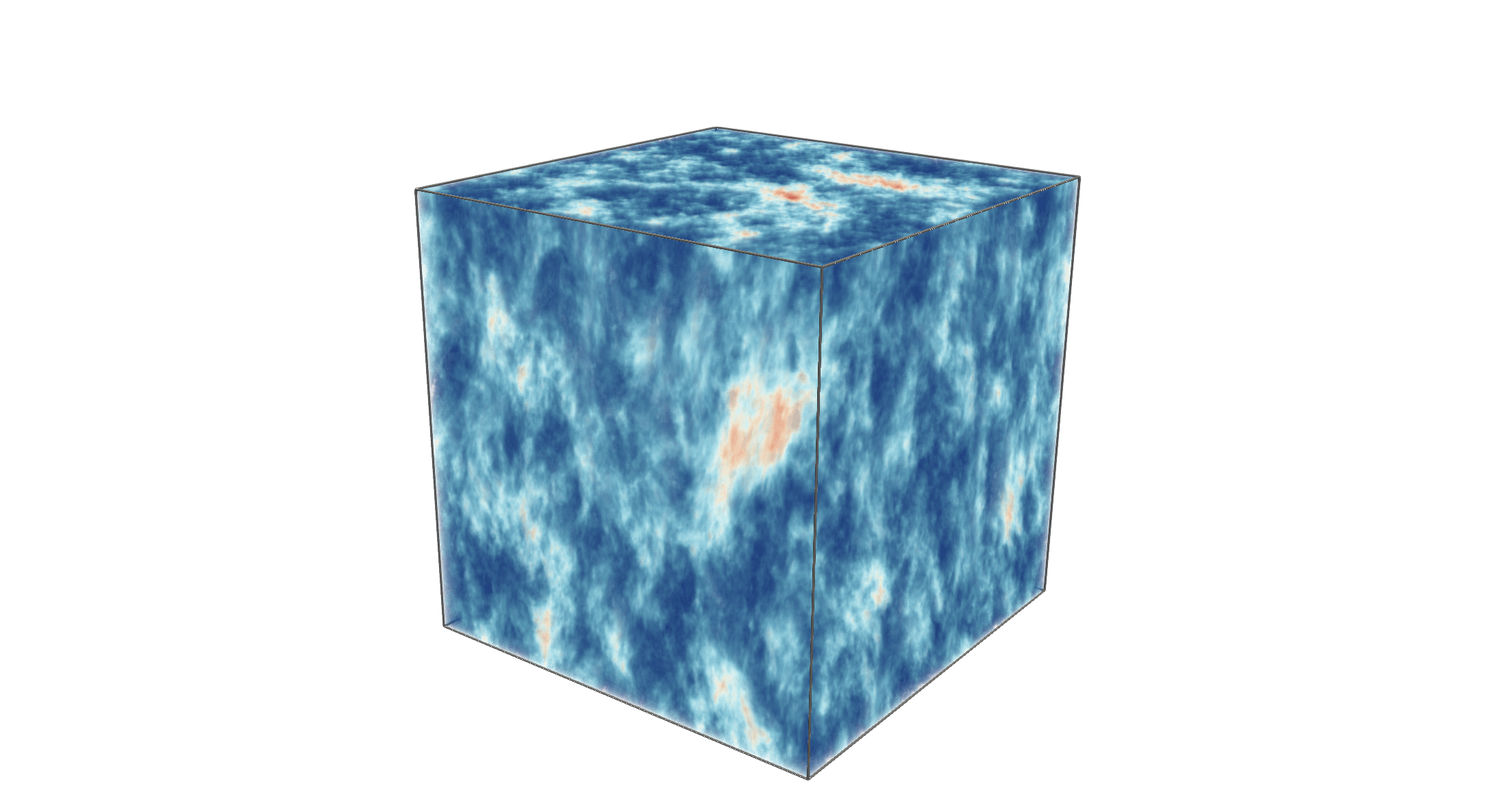}
	~
	\includegraphics[clip=true, trim = 21cm 1cm 22cm 6cm, width=0.28\textwidth]{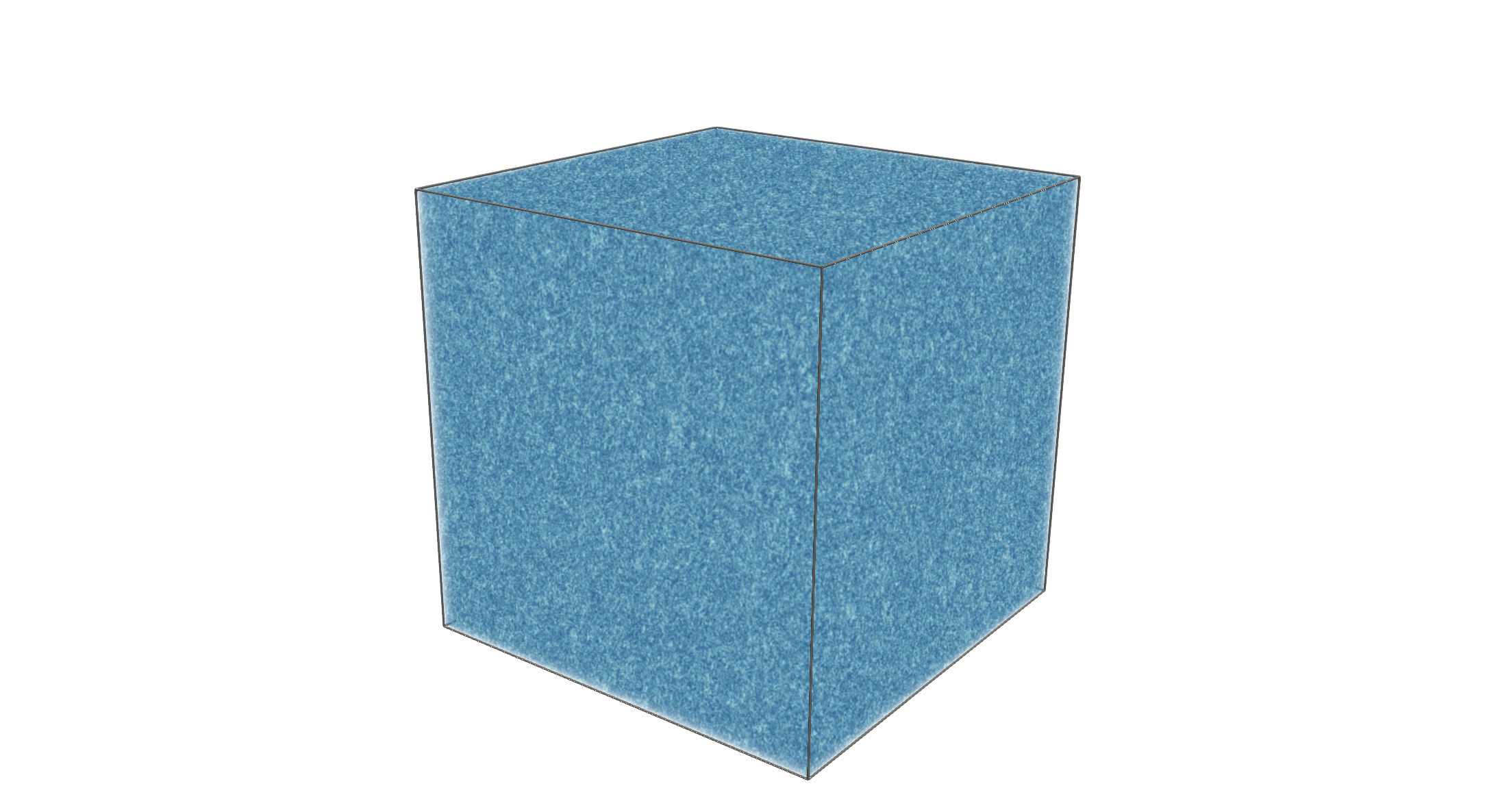}
	\caption{\label{fig:PotentialVelocityVorticity}Normalized magnitudes of $\bpsi$ (left), $\bfu = \curl\bpsi$ (center), and $\bfw = -\Delta\bpsi$ (right). 
		Observe the decrease of regularity, from left to right, with higher order derivatives of the vector potential.
	The fields are computed using a discrete Fourier transform.}
\end{figure}

\begin{remark}
	Note that the vector potential $\bpsi(\bmx)$, defined in~\cref{eq:VectorPotentialFourier}, is not divergence-free.
	In an alternative model, one may seek to enforce this condition.
	In this case, one would naturally arrive at the Stokes-type system
	\begin{equation}
	\label{eq:DivFreeCharacterizationOfPsi}
		A^{\alpha} \bpsi + \nabla \phi = \mu L^{2\alpha} \bxi
		,
		\qquad
		\div \bpsi = 0
		.
	\end{equation}
	Here, $\phi$ plays the role of an additional pressure-like Lagrange multiplier.
	Note that by taking the curl of the first equation above, the turbulence $\bfu(\bmx)$ can be characterized by just one equation; namely,
	\begin{equation}
	\label{eq:DirectCharacterizationOfu}
		A^{\alpha} \bfu = \mu L^{2\alpha} \curl \bxi
		\,.
	\end{equation}

	For the sake of completeness, note that we may also define a generalized vorticity field $\bfw = -\Delta \bpsi$.
	One may show that $\bfw(\bmx) = \frac{1}{\sqrt{4\upi}} \int E^{1/2}(k) \ee^{\ii\bmk\bcdot\bmx} \dd \bmW(\bmk)$.
	This expression, in combination with the PDE
	\begin{equation}
	\label{eq:VorticityCharacterizationOfu}
		-\Delta \bfu = \curl \bfw
		,
	\end{equation}
	can also be used to characterize $\bfu(\bmx)$.

	Both~\cref{eq:DirectCharacterizationOfu,eq:VorticityCharacterizationOfu} are perfectly valid and equivalent characterizations of the homogeneous turbulent velocity field considered above, $\bfu(\bmx)$, on the free space domain $\R^3$.
	More importantly, they will likely lead to alternative turbulence models on more complicated domains, once appropriate boundary conditions are selected.
	We have chosen not to use~\cref{eq:DirectCharacterizationOfu} because it is not valid in the presence of non-homogeneous length scales $L = L(\bmx)$; a modeling consideration we wish to incorporate.
	The non-homogeneous setting still requires the saddle-point problem~\cref{eq:DivFreeCharacterizationOfPsi} in order to enforce volume conservation in $\bpsi(\bmx)$.
	Because $\bfu = \curl \bpsi$ does not depend on the irrotational part of $\bpsi(\bmx)$,~\cref{eq:DivFreeCharacterizationOfPsi} appears to be a valid alternative model which we leave open for future investigation.
	Finally, we have chosen to avoid~\cref{eq:VorticityCharacterizationOfu} because of the low regularity of the solution variable $\bfw(\bmx)$; cf. \cref{fig:PotentialVelocityVorticity}.
\end{remark}

\section{Main results} %
\label{sec:general_models}

In this section, we relate a large class of turbulent vector fields $\bfu$ to the solution of a general family of FPDEs with stochastic forcing.
In particular, we put forth a general inhomogeneous model, derive a corresponding model for shear flows, and motivate a physically meaningful choice of boundary conditions.

\subsection{A general class of inhomogeneous models} %
\label{sub:general_model}

Equation~\cref{eq:VKFractionalPDE} was derived from a very specific form of the energy spectrum function $E(k)$.
Under the same decomposition of the spectral tensor $\Phi(\bmx)$ given in~\cref{eq:IsotropicSpectralTensor}, a much more general family of homogeneous and stationary random field models derive from the following ansatz on the energy spectrum function:
\begin{equation}
\label{eq:GeneralSpectrumAnsatz}
	k^{-4}E(\bmk)
	=
	\mu^2 \det(\bar{\bTheta})^{2/3\gamma}(1 + \bmk^\top\bar{\bTheta}\bmk)^{-2\alpha_1}(\bmk^\top\bar{\bTheta}\bmk)^{-2\alpha_2}
	.
\end{equation}
Here, $\bar{\bTheta} \in \R^{3\times 3}$ is a fixed symmetric positive definite matrix and $\alpha_2$, $\alpha_1$, $\gamma$, and $\mu$ are additional scalar parameters.

Just as $L$ played the role of a length scale in~\cref{eq:VKEnergySpectrum}, here, $\bar{\bTheta}$ plays the role of a metric in Fourier space.
Observe that if $\bar{\bTheta}=L^2\bsfI$, where $\bsfI$ denotes the identity matrix, $4\alpha_2=4-p_0$, $4\alpha_1=5/3+p_0$, $\gamma = \alpha_1+\alpha_2$, and $\mu^2=C \varepsilon^{2/3}$, then~\cref{eq:GeneralSpectrumAnsatz} reproduces the following common one-parameter homogeneous energy spectrum model \citep[see, e.g.,][p.~232]{pope2001turbulent}:
\begin{equation}
\label{eq:PaoSpectrum}
	E(k) =
	C \varepsilon^{2/3}k^{-5/3} \bigg(
	\frac{k L}{((k L)^2 + 1)^{1/2}}
	\bigg)^{5/3 + p_0}.
\end{equation}
Here, the scenario $p_0=4$ corresponds exactly to the \VonKarman{} spectrum~\eqref{eq:VKEnergySpectrum} considered previously; i.e., $\alpha_1 = \gamma = \frac{17}{12}$ and $\alpha_2 = 0$.

As in~\cref{eq:VectorPotentialFourier}, the vector potential~$\bpsi(\bmx) = \int \ee^{\ii\bmk\bcdot\bmx} \dd \bmY(\bmk)$ can also be written in terms of a Fourier--Stieltjes integral, weighted by $k^{-2} E^{1/2}(\bmk)$.
After rearranging factors,~\cref{eq:GeneralSpectrumAnsatz} characterizes the vector potential~$\pot$ as the solution of the following fractional stochastic PDE on $\R^3$:
\begin{equation}\label{eq:nearly_general_SPDE}
\big(\Id-\nabla\bcdot(\bar{\bTheta}\nabla)\big)^{\alpha_1}\big(-\nabla\bcdot(\bar{\bTheta}\nabla)\big)^{\alpha_2}\pot
=
\mu\det(\bar{\bTheta})^{\gamma/3}\bxi.
\end{equation}

Two immediate modifications of~\cref{eq:nearly_general_SPDE} are now in order.
First, we may replace the constant matrix $\bar\bTheta$ by a spatially varying metric tensor $\bTheta(\bmx)$.
This change immediately induces an inhomogeneous turbulence model.
Second, we may consider substituting the white noise random variable $\bxi$ for a well-chosen colored noise variable denoted $\bmeta$.
Together, these two generalizations lead to a family of random field models written
\begin{equation}\label{eq:general_SPDE}
	\big(\Id-\nabla\bcdot(\bTheta(\bmx)\nabla)\big)^{\alpha_1}\big(-\nabla\bcdot(\bTheta(\bmx)\nabla)\big)^{\alpha_2}\pot
	=
	\mu\det(\bTheta(\bmx))^{\gamma/3}\bmeta
	.
\end{equation}

Physically, the metric tensor~$\bTheta(\bmx)$ introduces inhomogeneous and anisotropic diffusion; this corresponds to local changes of the turbulence length scales which may result from complicated dynamics of interacting eddies.
Statistically, it incorporates the possibility for spatially varying correlation lengths and also may contain distortion.

In order to motivate one possible choice in the stochastic forcing term $\bmeta$, note that~\cref{eq:PaoSpectrum} can adequately characterize both the energy-containing and inertial subranges, however, it fails in the dissipative range; namely, where $k$ is large.
In order to fit the dissipative range, one approach is to define the energy spectrum as the product of~\cref{eq:PaoSpectrum} and a decaying exponential function like
\begin{equation*}
	f_{\beta}(k)
	=
	\ee^{-\beta k}
	,
\end{equation*}
where $\beta>0$ is a positive constant, usually close to the Kolmogorov length scale. 
In such scenarios, we suggest using the following definition for $\bmeta$ in~\cref{eq:general_SPDE}:
\begin{equation*}
\label{eq:PaoNoise}
	\bxi_\beta(\bmx)
	=
	\int_{\R^3}
	\ee^{\ii\bmk\bcdot\bmx} f_{\beta}(k) \dd \bmW(\bmk)
	\propto
	\bxi(\bmx)\ast\frac{\beta}{\beta^2 + |\bmx|^2}
	,
\end{equation*}
which converges to~\eqref{eq:ksi} as $\beta\to 0$.
In the presence of shear, a different time-dependent modification is also natural to consider from the point of view of rapid distortion theory.
That is the subject of the following subsection.

\begin{remark}
	When $\alpha_2$ and $\alpha_1$ are chosen to match the energy spectrum model~\cref{eq:PaoSpectrum}, it is clear that $\alpha_2+\alpha_1 = 17/12$ is independent of~$p_0$.
	Under this constraint, $\alpha_2$ and $\alpha_1$ mainly affect the behavior of the power spectrum at the origin and, likewise, the large scale structure of $\bfu$.
	In other words, the shape of the spectrum in the inertial subrange is unaffected by the precise choice of $\alpha_2$ and $\alpha_1 = 17/12-\alpha_2$; only the shape of the spectrum in the energy-containing range is affected.
\end{remark}

\subsection{A simple instationary model for shear flows} %
\label{sub:rapid_distortion}

Consider the velocity field $\bfU = \langle\bfU\rangle + \bfu$ and define the average total derivative of the turbulent fluctuations $\bfu = (u_1,u_2,u_3)$ as follows:
\begin{equation*}
	\frac{\bar{D} u_i}{\bar{D} t}
	=
	\frac{\partial u_i}{\partial t}
	+
	\langle U_j\rangle\frac{\partial u_i}{\partial x_j}
	\,.
\end{equation*}
The rapid distortion equations~\citep[see, e.g.,][]{townsend1980structure,Maxey1982,Hunt1990} are a linearization of the Navier--Stokes equations in free space when the turbulence-to-mean-shear time scale ratio is arbitrarily large.
They can be written
\begin{equation}
\label{eq:RapidDistortionEquations}
	\frac{\bar{D} u_i}{\bar{D} t}
	=
	-u_i \frac{\partial \langle U_j\rangle}{\partial x_i} - \frac{1}{\rho}\frac{\partial p}{\partial x_i}
	,
	\qquad
	\frac{1}{\rho} \Delta p
	=
	-2\frac{\partial \langle U_i\rangle}{\partial x_j}\frac{\partial u_j}{\partial x_i}
	\,.
\end{equation}

Under a uniform shear mean velocity gradient, $\langle U_i(\bmx)\rangle = x_j\partial \langle U_i\rangle / \partial x_j$, where $\partial \langle U_i\rangle / \partial x_j$ is a constant tensor, a well-known form of these equations can be written out in Fourier space.
In this case, the rate of change of each frequency $\bmk(t) = (k_1(t),k_2(t),k_3(t))$ is defined ${\!\dd k_i}/{\!\dd t} = -k_j{\partial \langle U_j\rangle}/{\partial x_i}$.
We then have the following Fourier representation of the average total derivative of $\bfu$:
\begin{equation*}
	\frac{\bar{D} u_i}{\bar{D} t}
	=
	\int_{\R^3}
	\ee^{\ii\bmk\bcdot\bmx}\Bigg(
		\bigg(\frac{\partial }{\partial t} + \frac{\dd k_j}{\dd t}\frac{\partial }{\partial k_j}\bigg) \dd Z_i(\bmk,t)
	\Bigg)
	=
	\int_{\R^3}
	\ee^{\ii\bmk\bcdot\bmx}\Bigg(\frac{\bar{D} \dd Z_i(\bmk,t)}{\bar{D} t}\Bigg)
	.
\end{equation*}
With this expression, the Fourier representation of~\cref{eq:RapidDistortionEquations} can be written
\begin{equation}
\label{eq:RapidDistortionEquationsFourier}
	\frac{\bar{D} \dd Z_j(\bmk,t)}{\bar{D} t}
	=
	\frac{\partial U_\ell}{\partial x_k}
	\bigg(
		2 \frac{k_j k_\ell}{k^2} - \delta_{j\ell}
	\bigg)
	\dd Z_k(\bmk,t)
	\,.
\end{equation}

Exact solutions to~\cref{eq:RapidDistortionEquationsFourier} are well-known \citep[see, e.g.,][]{townsend1980structure,mann1994spatial}, given the initial conditions $\bmk_0 = (k_{10},k_{20},k_{30})$ and $\!\dd \bfZ(\bmk_0,0)$.
In the scenario
\begin{equation*}
	\langle \bfU(\bmx)\rangle
	=
	(U_0 + S x_3) \bme_1
	,
\end{equation*}
the solution can be written in terms of the evolving Fourier modes $\bmk(t)$ and non-dimensional time $\tau = S t$, as follows:
\begin{equation*}
	\dd \bfZ(\bmk,t)
	=
	\bsfD_\tau(\bmk)
	\dd \bfZ(\bmk_0,0)
	,
\end{equation*}
where
\begin{equation*}
	\bsfD_\tau(\bmk)
	=
	\begin{bmatrix*}
		1 & 0 & \zeta_1\\
		0 & 1 & \zeta_2\\
		0 & 0 & \zeta_3
	\end{bmatrix*}
	,
	\qquad
	\bmk_0 = \bsfT_\tau \bmk
	,
	\qquad
	\bsfT_\tau = 
	\begin{bmatrix}
	1 & 0 & 0 \\
	0 & 1 & 0 \\
	\tau & 0 & 1
	\end{bmatrix}
	.
\end{equation*}
In the expression for $\bsfD_\tau(\bmk)$, the non-dimensional coefficients $\zeta_i = \zeta_i(\bmk,\tau)$, $i=1,2,3$, are defined
\begin{equation*}
	\zeta_1 = C_1 - C_2 k_2/k_1,
	\quad
	\zeta_2 = C_1 k_2/k_1 + C_2,
	\quad
	\zeta_3 = k_0^2/k^2,
\end{equation*}
where $k_0 = |\bmk_0|$ and
\begin{equation*}
	C_1 = \frac{\tau k_1^2(k_0^2 -2k_{30}^2 + \tau k_1 k_{30})}{k^2(k_1^2+k_2^2)},
	\qquad
	C_2 = \frac{k_2 k_0^2}{(k_1^2+k_2^2)^{3/2}}
	\arctan\left(
	\frac{\tau k_1(k_1^2 + k_2^2)^{1/2}}{k_0^2 - \tau k_{30}k_1}
	\right)
	.
\end{equation*}

One may observe that
\begin{equation*}
	\begin{bmatrix*}
		1 & 0 & \zeta_1\\
		0 & 1 & \zeta_2\\
		0 & 0 & \zeta_3
	\end{bmatrix*}
	\begin{bmatrix*}
		0 & -k_{30} & k_2\\
		k_{30} & 0 & -k_1\\
		-k_2 & k_1 & 0
	\end{bmatrix*}
	=
	\begin{bmatrix*}
		0 & -k_{3} & k_2\\
		k_3 & 0 & -k_1\\
		-k_2 & k_1 & 0
	\end{bmatrix*}
	\begin{bmatrix*}
		\zeta_3 & 0 & 0\\
		0 & \zeta_3 & 0\\
		-\zeta_1 & -\zeta_2 & 1
	\end{bmatrix*}
	,
\end{equation*}
or, equivalently, $\bsfD_\tau(\bmk) k_0^{-2}\bsfQ(\bmk_0) = k^{-2}\bsfQ(\bmk) \bsfD_\tau^{-\top}(\bmk)$.
Moreover, $\dd\bmW(\bmk_0) = \dd\bmW(\bmk)$, due to translational invariance.
Therefore, taking
$
	\!\dd \bmZ(\bmk_0,0)
	=
	\bsfQ(\bmk_0) \Big(\frac{1}{\sqrt{4\upi}k_0^2} E^{1/2}(\bmk_0) \dd\bmW(\bmk_0)\Big)
	,
$
it holds that
\begin{equation*}
	\dd \bfZ(\bmk,t)
	=
	\bsfQ(\bmk) \bigg(\frac{1}{\sqrt{4\upi}k^2} E^{1/2}(\bsfT_\tau\bmk)\, \bsfD_\tau^{-\top}(\bmk)\dd\bmW(\bmk)\bigg)
	.
\end{equation*}
Finally, invoking the general expression for $E(\bmk)$ written in~\cref{eq:GeneralSpectrumAnsatz}, one arrives at the rapid distortion equation fractional PDE
\begin{equation}
\label{eq:RapidDistortionHomogeneousFPDE}
	\big(\Id-\nabla\bcdot(\bar{\bTheta}_\tau\nabla)\big)^{\alpha_1}\big(-\nabla\bcdot(\bar{\bTheta}_\tau\nabla)\big)^{\alpha_2}\pot
	=
	\mu\det(\bar{\bTheta}_\tau)^{\gamma/3}\bmeta_\tau
\end{equation}
where $\bar{\bTheta}_\tau = \bsfT_\tau^\top \bar{\bTheta} \bsfT_\tau$ and $\bmeta_\tau(\bmx) = \int_{\R^3} \ee^{\ii\bmk\bcdot\bmx} \big(\bsfD_\tau^{-\top}(\bmk)\dd\bmW(\bmk) \big)$.
Note that $\det(\bar{\bTheta}_\tau) = \det(\bar{\bTheta})$.

\begin{remark}
For each fixed $t$,~\cref{eq:RapidDistortionHomogeneousFPDE} is clearly a particular case of~\cref{eq:general_SPDE}.
The generalization of this model to an inhomogeneous instationary FPDE is discussed in \cref{sub:shear_boundary_layers}.
\end{remark}

\begin{remark}
	An important extension of the rapid distortion model above involves replacing the constant $\tau$ by a wavenumber-dependent ``eddy lifetime'' $\tau(k)$; see, e.g.,~\citet{mann1994spatial}.
	Such models are considered more realistic because, at some point, the shear from the mean velocity gradient will cause the eddies to stretch and eventually they will breakup within a size-dependent timescale.
	In this case, the generalization of $\bmeta_\tau$ above is straightforward.
	Meanwhile, at least when~$\overline{\bTheta} = L^2 \bsfI$, one may consider replacing the operator $\overline{\bTheta}_\tau$ in~\cref{eq:RapidDistortionHomogeneousFPDE} by
	\begin{equation*}\label{key}
		L^2\,
		\Fourier\inv
		\begin{bmatrix}
		1+\tau(k)^2 & 0 & \tau(k) \\
		0 & 1 & 0 \\
		\tau(k) & 0 & 1
		\end{bmatrix}
		\Fourier.
	\end{equation*}
	To solve such an equation numerically, one doesn't need to construct the closed form of the linear operator, but may instead choose to use a matrix-free Krylov method \citep{saad2003iterative}.
\end{remark}

\subsection{Boundary conditions} %
\label{sub:boundary_conditions}

There are a number of different, equivalent, definitions of fractional operators on $\R^3$.
However, moving from the free-space equation~\cref{eq:general_SPDE} to a boundary value problem relies on heuristics and can be done in a wide variety of ways; each of which may also differ by the specific definition of the fractional operator being used \citep{lischke2020fractional}.
As stated previously, in this work, we choose to only deal with the spectral definition.
In this setting, boundary conditions are applied to the corresponding integer-order operator and then incorporated implicitly by modifying the spectrum; cf.~\cref{eq:AbstractSpectralRepresentation,eq:SpectramTheorem}.

Assume that~\cref{eq:general_SPDE} is posed on a three-dimensional simply-connected domain $\Omega\subsetneq \R^3$ with boundary $\Gamma = \bdry\Omega$.
We begin with the following heuristically chosen impermeability condition for the velocity field:
\begin{equation}
	\bfu = \curl \bpsi
	\quad
	\text{in } \Omega
	,
	\qquad
	\bfu\bcdot\bmn = 0
	\quad
	\text{on } \Gamma
	.
\label{eq:Non-penetrationCondition}
\end{equation}
Although more relaxed boundary conditions are of course also possible, we choose to enforce~\cref{eq:Non-penetrationCondition} via a no-slip condition on the vector potential $\bpsi$; specifically,
\begin{equation}
	\bpsi - (\bpsi\bcdot\bmn)\bmn
	=
	\bzero
	\quad
	\text{on }
	\Gamma
	.
\label{eq:NoSlipPotential}
\end{equation}
The remaining boundary condition must restrict $\bpsi$ normal to $\Gamma$ and is, therefore, independent of the requirement $\curl \bpsi\bcdot \bmn = 0$.
One natural choice is the generalized (homogeneous) Robin condition
\begin{equation}
	\kappa\sspace\bpsi\bcdot\bmn + (\bTheta(\bmx)\nabla\bpsi)\bmn\bcdot \bmn
	=
	0
	\quad
	\text{on }
	\Gamma
	.
\label{eq:RobinPotential}
\end{equation}
Here, the new model parameter $\kappa \geq 0$ can be inferred from available data.
Note that in the limit $\kappa \to \infty$, we uncover the impermeability boundary condition $\bpsi\bcdot\bmn = 0$.
Together with~\cref{eq:NoSlipPotential}, it implies the complete Dirichlet boundary condition, $\bpsi = \bzero$ on $\Gamma$.
Hereon, we use the notation $\kappa=\infty$ to indicate this limiting scenario.

Note that~\cref{eq:general_SPDE} can be written $\mcL\bpsi = \bmb$, where
\begin{equation*}
	\mcL
	:=
	\big(\Id-\nabla\bcdot(\bTheta(\bmx)\nabla)\big)^{\alpha_1}\big(-\nabla\bcdot(\bTheta(\bmx)\nabla)\big)^{\alpha_2}
	\quad\text{and}\quad
	\bmb := \mu\det(\bTheta(\bmx))^{\gamma/3}\bmeta
	.
\end{equation*}
In order to define the domain $\mcD(\mcL)$ of the multi-fractional operator $\mcL\colon \mcD(\mcL) \subset [L^2(\Omega)]^3 \to [L^2(\Omega)]^3$, we start by letting $A := \big(\Id-\nabla\bcdot(\bTheta(\bmx)\nabla)\big) \colon \mcD(A) \subset [L^2(\Omega)]^3 \to [L^2(\Omega)]^3$.
For notational convenience, we assume that $A$ has a discrete spectrum.

In the spectral definition of $A^{\alpha_1}$, the domain $\mcD(A)$ characterizes the boundary conditions on $\Gamma$.
In this work, assuming that $\det(\bTheta(\bmx))$ is uniformly bounded from above and below by positive constants, we define
\begin{equation*}
	\mcD(A)
	=
	\big\{
		\bpsi \in [H^2(\Omega)]^3\, \colon \text{\cref{eq:NoSlipPotential} and \cref{eq:RobinPotential} hold in the sense of traces}\,
	\big\}
	.
\end{equation*}
For this operator domain, there exists an orthonormal basis of eigenvectors $\{\bma_j\}_{j=1}^\infty\subset \mcD(A)$, with corresponding eigenvalues $\{a_j\}_{j=1}^\infty$ in non-increasing order; cf. \citet{bolin2020numerical}.
Then, following~\cref{eq:SpectramTheorem}, the fractional differential operator $A^{\alpha_1} \colon \mcD(A^{\alpha_1}) \subset [L^2(\Omega)]^3 \to [L^2(\Omega)]^3$ is defined
\begin{equation*}
	A^{\alpha_1}\bpsi
	=
	\sum_{j=1}^\infty
	a_j^{\alpha_1} (\bpsi,\bma_j)_\Omega\, \bma_j
\end{equation*}
and $\mcD(A^{\alpha_1}) = \{ \bpsi \in [L^2(\Omega)]^3\, \colon \sum_{j=1}^\infty a_j^{2\alpha_1} \sspace (\bpsi,\bma_j)_\Omega^2 < \infty\}$.

Now consider $A-\Id \colon \mcD(A) \to [L^2(\Omega)]^3$ and note that $\mcL = A^{\alpha_1} (A-\Id)^{\alpha_2}$. 
In this case, $A^{\alpha_1}$ and $(A-\Id)^{\alpha_2}$ commute because they share the same eigenmodes:
\begin{equation*}
	A^{\alpha_1} (A-\Id)^{\alpha_2} \bpsi
	=
	\sum_{j=1}^\infty
	a_j^{\alpha_1} (a_j-1)^{\alpha_2} (\bpsi,\bma_j)_\Omega\, \bma_j
	=
	(A-\Id)^{\alpha_2} A^{\alpha_1}\bpsi
	.
\end{equation*}
Accordingly, we define the domain of the operator $\mcL$ as follows:
\begin{equation}
\label{eq:DefinitionOfDomain}
	\mcD(\mcL)
	=
	\Bigg\{ \bpsi \in [L^2(\Omega)]^3\, \colon \sum_{j=1}^\infty a_j^{2\alpha_1}(a_j-1)^{2\alpha_2} \sspace (\bpsi,\bma_j)_\Omega^2 < \infty\Bigg\}
	.
\end{equation}

We may now write the boundary value problem given by~\cref{eq:general_SPDE,eq:NoSlipPotential,eq:RobinPotential} as the abstract operator equation $\mcL\bpsi = \bmb$, with $\mcD(\mcL)$ defined in~\cref{eq:DefinitionOfDomain}.
Nevertheless, we will still usually refer to this problem in the ``strong form''
\begin{equation}
\label{eq:BVP}
\left\{
\begin{alignedat}{3}
	\big(\Id-\nabla\bcdot(\bTheta(\bmx)\nabla)\big)^{\alpha_1}\big(-\nabla\bcdot(\bTheta(\bmx)\nabla)\big)^{\alpha_2}\bpsi
	&=
	\mu\det(\bTheta(\bmx))^{\gamma/3}\bmeta
	\quad
	&&\text{in } \Omega,
	\\
	\bpsi - (\bpsi\bcdot\bmn)\bmn
	&= 0
	\quad
	&&\text{on } \Gamma,
	\\
	\kappa\sspace\bpsi\bcdot\bmn
	+
	(\bTheta(\bmx)\nabla\bpsi)\bmn\bcdot\bmn
	&= 0
	\quad
	&&\text{on } \Gamma,
\end{alignedat}
\right.
\end{equation}
since it is much more physically illustrative.

\section{Physical applications} %
\label{sec:applications}

In this section, we document three applications of~\cref{eq:BVP} and some theoretical results.
The first two applications describe turbulent conditions which may be modeled using the general FPDE model~\cref{eq:BVP}.
In the final subsection, we highlight an important wind engineering application.
Here, the model is used to generate a turbulent inlet profile for a numerical wind tunnel simulation of the atmospheric boundary layer.

\subsection{Shear-free boundary layers} %
\label{sub:shear_free_boundary_layers}

There are many different examples of turbulence confined by a solid boundary, without any significant mean shear \citep{Hunt1984turbulence}.
In such flows, the rate of turbulent kinetic energy dissipation $\epsilon$ can be assumed to be {\it approximately} constant with height.
This setting has been studied in detail by various authors (see, e.g., \citet{Hunt1984turbulence,Hunt1989Cross,Perot1995,perot1995shear,Aronson1997} and references therein) and so forms a solid proving ground to validate~\cref{eq:BVP}.

\subsubsection{A \VonKarman{}-type model} %
\label{ssub:the_von_karman_spectrum}

We begin with the inhomogeneous turbulence model \cref{eq:BVP}, with fractional coefficients corresponding to the \VonKarman{} energy spectrum~\cref{eq:VKEnergySpectrum}, on the open half space domain $\R_+^3 = \{ (x,y,z) \in\R^3 \colon z >0 \}$.
Based on the supposed absence of shear, we also consider the following simple diagonal form for the diffusion tensor, in Cartesian coordinates:
\begin{equation*}\label{eq:SFBL_Theta}
	\ten{\Theta}(z)
	=
	\begin{bmatrix}
		L_{1}(z)^2 & 0 & 0 \\
		0 & L_{2}(z)^2 & 0 \\
		0 & 0 & L_{3}(z)^2
	\end{bmatrix}
	\,.
\end{equation*}
Defining $L(z) = \sqrt[3]{L_1(z)L_2(z)L_3(z)}$, the appropriate form of~\cref{eq:BVP} can be written as follows:
\begin{equation}
\label{eq:VKFractionalPDE_BVP_halfspace}
\left\{
\begin{alignedat}{3}
	\big(\Id-\div(\ten{\Theta}(z)\grad)\big)^{17/12} \bpsi
	&=
	\mu\sspace L(z)^{17/6}\sspace \bxi
	\quad
	&&\text{in } \R_+^3,
	\\
	\kappa\sspace\psi_3 + L_{3}(z)^2 \frac{\partial\psi_3}{\partial z}
	=
	\psi_1 &= \psi_2 =
	0
	\quad
	&&\text{at } z=0
	.
\end{alignedat}
\right.
\end{equation}

\begin{figure}
\centering
	\includegraphics[width=0.65\textwidth]{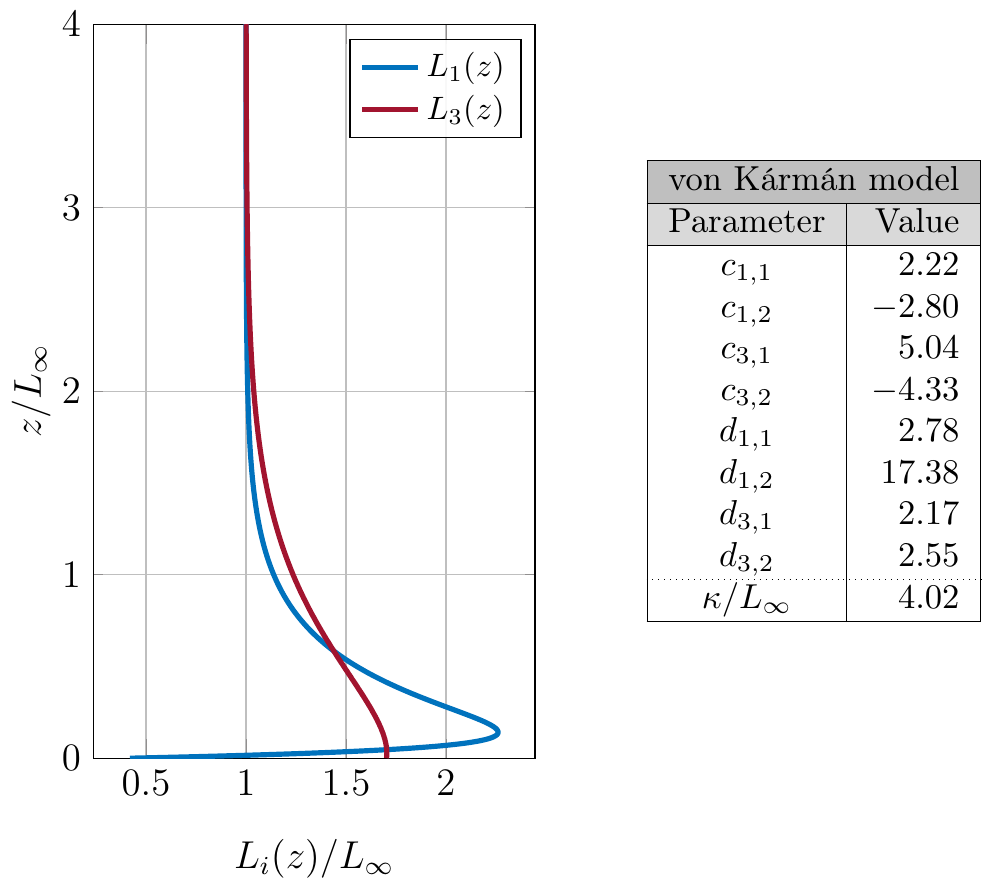}
	\caption{\label{fig:L(z)}Optimal diffusion coefficients $L_i(z)$ and Robin constant $\kappa$ determined by fitting the Reynolds stress data in \cref{fig:ThomasHancockData}. Note that $L_1(z) = L_2(z)$.}
\end{figure}

Both the Robin coefficient~$\kappa$ and an explicit parametric expression for each $L_i(z)$ give rise to a model design parameter vector, say $\btheta$.
This vector $\btheta$ may then be subject to calibration with respect to experimental data, e.g., using the technique described in~\cref{sub:fitting_data}.
This process of model calibration is important because wall roughness, Reynolds number, and the nature of the turbulence may affect the near-wall statistics \citep{pope2001turbulent} and may be incorporated through proper parameter selection.
For instance, let us consider the following exponential expansion
\begin{equation}\label{eq:Expansion}
	L_i(z)
	=
	L_\infty \cdot\bigg(1 + \sum_{k=1}^K c_{i,k}\ee^{-d_{i,k}\sspace \frac{z}{L_\infty}} \bigg)
	,
\end{equation}
with each $d_{i,k}\ge0$, $c_{1,k} = c_{2,k}$ and $d_{1,k} = d_{2,k}$.
Taking only two terms in each expansion above ($K=2$), we arrive through calibration at a statistical model which closely matches the experimental data found in~\citet{Thomas1977}.
Note that with such a model, $L_1(z) = L_2(z)$ and each $L_i(z)$ exponentially converges to the homogeneous length scale $L_\infty$, as $z\to \infty$, as illustrated in~\cref{fig:L(z)}.

The prescribed boundary conditions will affect the physical length scales of the random velocity field $\bfu = \nabla \times \bpsi$.
Therefore, the diffusion coefficients~$L_i(z)$ do not necessarily correspond to the physical length scales.
For this reason, we follow~\citet{Lee1991} and define the (physical) so-called integral length scales
\begin{equation*}
	\ell_{ij}^{(x_m)}(z)
	=
	\frac{\int_{\R}\langle u_i(\bmx + r\vec{e}_m)\sspace u_j(\bmx)\rangle\dd r}{\langle u_i(\bmx)\sspace u_j(\bmx)\rangle}
	=
	\frac{\int_{\R} R_{ij}(r\vec{e}_m,z) \dd r}{R_{ij}(\bzero,z)}
	.
\end{equation*}
In the expressions above, we have accounted for the fact that all solutions of~\cref{eq:VKFractionalPDE_BVP_halfspace} are temporary stationary and statistically homogeneous in the $x$- and $y$-directions; i.e., $R(\bmr,\bmx,t) = R(\bmr,z)$.

\begin{figure}
\centering
	\includegraphics[width=0.45\textwidth]{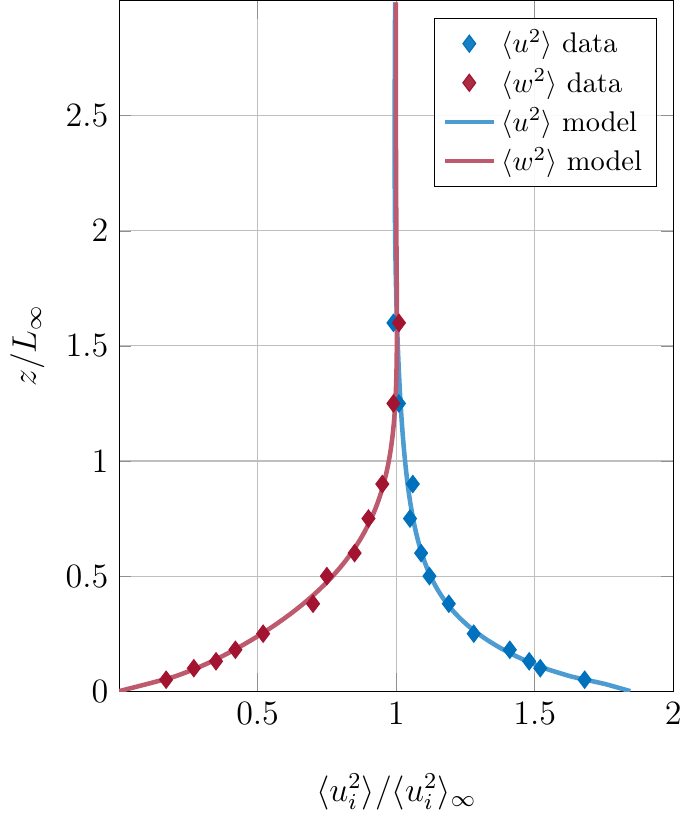}
	\quad
	\includegraphics[width=0.45\textwidth]{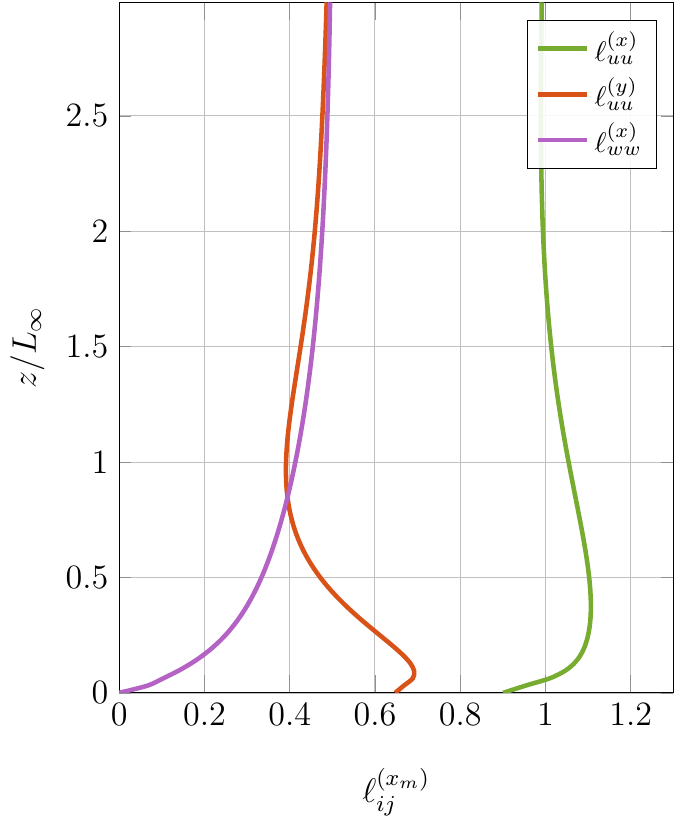}
	\caption{\label{fig:ThomasHancockData}Reynolds stress data from \cite{Thomas1977} compared with Reynolds stresses from the calibrated SFBL turbulence model~\cref{eq:VKFractionalPDE_BVP_halfspace} (left) and corresponding integral length scales (right).
	Observe that the model is able to closely fit the experimental data.}
\end{figure}

In~\cref{sec:calibration_and_rational_approximation}, we explain how to solve this problem numerically and to calibrate its solutions to Reynolds stress data.
The difference between the Reynolds stress profiles in the calibrated model and the corresponding experimental data is depicted in~\cref{fig:ThomasHancockData}, alongside the resulting integral length scales $\ell_{ij}^{(x_m)}(z)$.
Because this model has many free parameters which can be calibrated to experimental data, it is much more flexible than the classical theory proposed by Hunt et al.
Indeed, a comparison between the two theories, which highlights this flexibility, is given in the next subsection.
Note that the exact definitions of the optimized model parameters used in the results above are stated explicitly in the table in~\cref{fig:L(z)}.

\subsubsection{Comparison to the classical theory} %
\label{ssub:comparison_to_classical_theory}

It is important to consider the special case of~\cref{eq:VKFractionalPDE_BVP_halfspace} where each $L_i(z)$ is constant in $z$.
In Hunt's idealized SFBL theory \citep{Hunt1978free,Hunt1984turbulence}, derived from the energy spectrum ansatz~\cref{eq:VKEnergySpectrum} and briefly summarized in~\cref{sec:relationship_to_previous_work}, one can show that
\begin{equation*}
	\frac{\langle u^2 \rangle}{\langle u^2_\infty \rangle} = \frac{\langle v^2 \rangle}{\langle v^2_\infty \rangle} \to 1.5
	\quad\text{and}
	\quad
	\frac{\langle w^2 \rangle}{\langle w^2_\infty \rangle} =  O\bigg(\Big(\frac{z}{L_\infty}\Big)^{2/3}\bigg)
	\qquad\text{as}
	\quad
	\frac{z}{L_\infty} \to 0
	\,,
\end{equation*}
where $\langle u^2_\infty \rangle = \langle v^2_\infty \rangle = \langle w^2_\infty \rangle$ denotes the far field limit $z\to\infty$ of the non-zero Reynolds stresses.
The limit $\langle u^2 \rangle/\langle u^2_\infty \rangle \to 1.5$ is not always achieved in experiments (cf. \cref{fig:ThomasHancockData}), however, the limiting behavior $\langle w^2 \rangle/\langle w^2_\infty \rangle = O\big((z/L_\infty)^{2/3}\big)$ is well-established in the literature \citep{priestley1959turbulent,kaimal1976turbulence}.

\begin{figure}
\centering
	\includegraphics[width=0.45\textwidth]{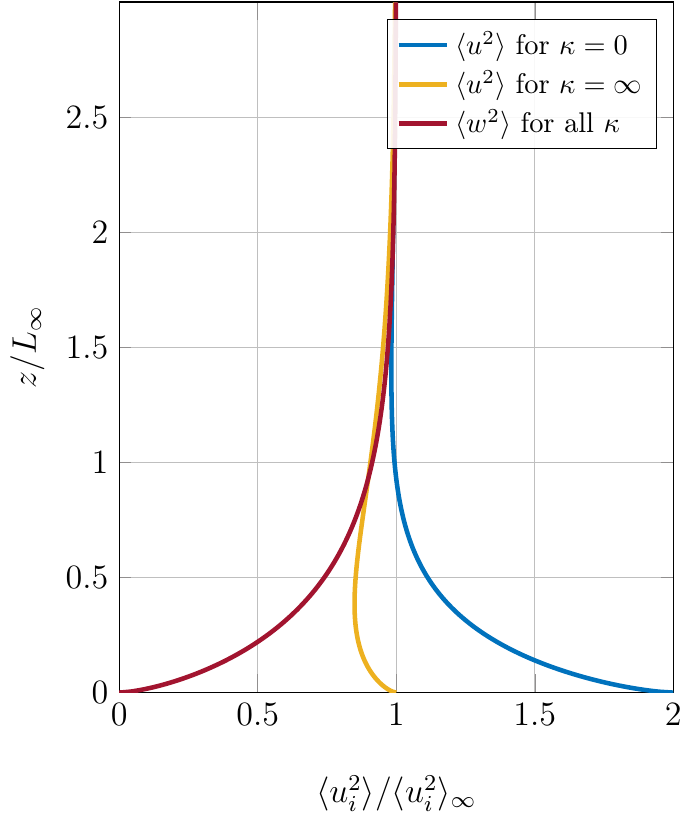}
	\quad
	\includegraphics[width=0.45\textwidth]{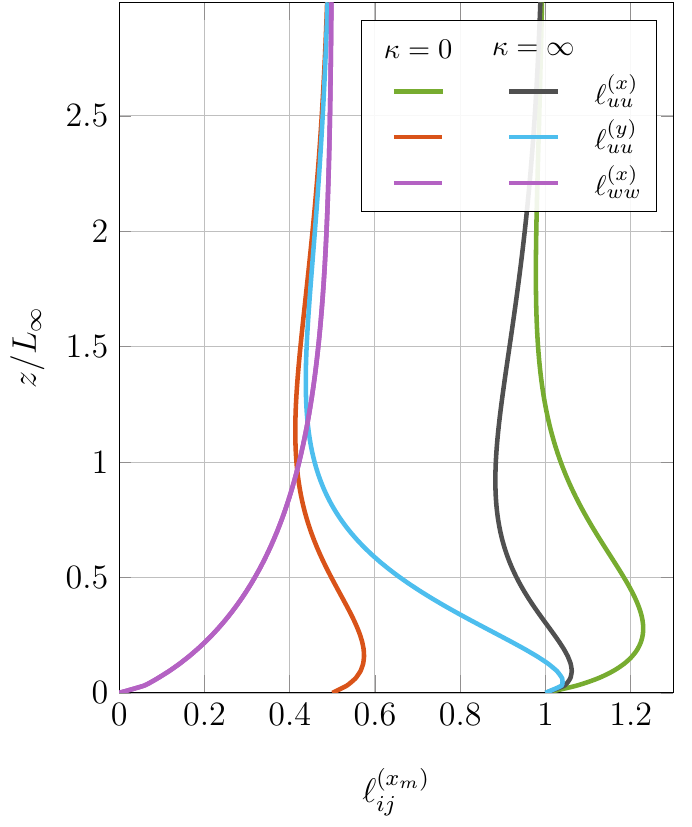}
	\caption{\label{fig:ReynoldsStressTheory}The analytically derived nonzero Reynolds stresses stated in~\cref{lem:Exact_soln_ww,lem:Exact_soln_uu_N,lem:Exact_soln_uu_D} (left) and corresponding integral length scales (right).}
\end{figure}

The corresponding scenario in our class of models is exactly~\cref{eq:VKFractionalPDE_BVP_halfspace} with each $L_i = L_\infty$.
In this setting, the nonzero Reynolds stresses, $\langle u^2 \rangle = \langle v^2 \rangle$ and $\langle w^2 \rangle$, can be derived analytically, at least for certain values of $\kappa\geq 0$.
These exact analytical solutions are summarized in~\cref{lem:Exact_soln_ww,lem:Exact_soln_uu_N,lem:Exact_soln_uu_D}, the proofs of which can be found in~\cref{app:proofs}.
Exact analytical solutions for the integral length scales $\ell_{ij}^{(x_m)}(z)$ can also be derived by a similar technique, but we do not include their derivation in this work for the sake of brevity.
Plots of the analytical Reynolds stresses and integral length scales are depicted in \cref{fig:ReynoldsStressTheory}.

\begin{lemma}
\label{lem:Exact_soln_ww}
	Given $\bfu = (u,v,w) = \nabla\times\bpsi$, where $\bpsi$ is any solution of~\cref{eq:VKFractionalPDE_BVP_halfspace} with constant $L_1 = L_2 = L_\infty$,
	it holds that
	\begin{equation}\label{eq:lemma:statement}
	\frac{\avg{w^2}}{\avg{w^2_\infty}}
	=
	1 - \Matern_{1/3}\left(\frac{2z}{L_\infty}\right)
	,
	\end{equation}
	where $\Matern_{\nu}(x)$ is the Mat\'ern kernel~\citep{Matern1986,Stein1999,Khristenko2019} given by
	\begin{equation*}\label{key}
	\Matern_{\nu}(x) = \frac{x^\nu\BesselK_{\nu}(x)}{2^{\nu-1}\Gamma(\nu)},
	\qquad \nu \geq 0,
	\end{equation*}
	and $\BesselK_{\nu}(x)$ denotes the modified Bessel function of the second kind~\citep{abramowitz1948handbook,bateman1953higher,watson1995treatise}.
Moreover, near the boundary the following expansion holds:
	\begin{equation*}\label{key}
	\frac{\avg{w^2}}{\avg{w^2_\infty}}
	\sim
	\frac{\Gamma(2/3)}{\Gamma(4/3)} \left(\frac{z}{L_\infty}\right)^{2/3}
	\qquad\text{as}
	\quad
	\frac{z}{L_\infty} \to 0
	.
	\end{equation*}
\end{lemma}

\begin{lemma}
	\label{lem:Exact_soln_uu_N}
	Given $\bfu = (u,v,w) = \nabla\times\bpsi$, where $\bpsi$ is the solution of~\cref{eq:VKFractionalPDE_BVP_halfspace} with constant $\ten{\Theta} = L_\infty^2\bsfI$ and $\kappa=0$,
	it holds that
	\begin{equation*}%
	\frac{\avg{u^2}}{\avg{u^2_\infty}} = \frac{\avg{v^2}}{\avg{v^2_\infty}}
	=
	1 + (\nu+1)\Matern_{\nu}\left(\frac{2z}{L_\infty}\right) - \nu\Matern_{\nu+1}\left(\frac{2z}{L_\infty}\right)
	.
	\end{equation*}
	Hence, near the boundary,
	$
	\frac{\avg{u^2}}{\avg{u^2_\infty}} = \frac{\avg{v^2}}{\avg{v^2_\infty}} \to 2
	$
	as
	$\frac{z}{L_\infty} \to 0
	$.
\end{lemma}

\begin{lemma}
	\label{lem:Exact_soln_uu_D}
	Given $\bfu = (u,v,w) = \nabla\times\bpsi$, where $\bpsi$ is any solution of~\cref{eq:VKFractionalPDE_BVP_halfspace} with constant $\ten{\Theta} = L_\infty^2\bsfI$ and $\kappa=\infty$,
	it holds that
	\begin{equation*}%
		\frac{\avg{u^2}}{\avg{u^2_\infty}} = \frac{\avg{v^2}}{\avg{v^2_\infty}}
		=
		1 + \nu\Matern_{\nu}\left(\frac{2z}{L_\infty}\right) - \nu\Matern_{\nu+1}\left(\frac{2z}{L_\infty}\right)
		.
	\end{equation*}
	Hence, near the boundary,
	$
	\frac{\avg{u^2}}{\avg{u^2_\infty}} = \frac{\avg{v^2}}{\avg{v^2_\infty}} \to 1
	$
	as
	$
	\frac{z}{L_\infty} \to 0
	$.
\end{lemma}

\begin{remark}
	The Robin boundary condition $\kappa\sspace\psi_3 + L_\infty^2 \frac{\partial\psi_3}{\partial z} = 0$ has no effect on $\langle w^2 \rangle$.
	Therefore, the asymptotic expansion of the well-known \citep{priestley1959turbulent,kaimal1976turbulence,Hunt1984turbulence,Hunt1989Cross} asymptotic behavior $\langle w^2 \rangle/\langle w^2_\infty\rangle = O\big((z/L_\infty)^{2/3}\big)$ as $z/L_\infty \to 0$ always holds when $L_1 = L_2 = L_\infty$.
\end{remark}

\begin{remark}
The limit $\langle u^2 \rangle/\langle u^2_\infty \rangle \to 1.5$ from Hunt's theory lies exactly in between the range of analogous limits, $\langle u^2 \rangle/\langle u^2_\infty \rangle \to 1$ and $\langle u^2 \rangle/\langle u^2_\infty \rangle \to 2$, coming from the exact solutions of~\cref{eq:VKFractionalPDE_BVP_halfspace} when $\kappa = \infty$ and $\kappa = 0$, respectively.
Numerical experiments show that $\langle u^2 \rangle/\langle u^2_\infty \rangle = \langle v^2 \rangle/\langle v^2_\infty \rangle$ always limits to a value in the interval $(1,2)$ when $\kappa\in (0,\infty)$ and $\ten{\Theta} = L_\infty^2\bsfI$.
\end{remark}

\subsubsection{A more general energy spectrum} %
\label{ssub:a_more_general_spectrum_model}

In order to illustrate the dependence of~\cref{eq:BVP} on the parameter $\alpha_2 = 17/12 -\alpha_1$, we may consider an alternative form of~\cref{eq:VKFractionalPDE_BVP_halfspace} which corresponds to the energy spectrum~\cref{eq:PaoSpectrum} with $p_0 = 2$.
Here, for additional complexity, we also consider the load $\bmeta = \bxi_\beta$  with $\beta/L_\infty=10^{-2}$:
\begin{equation}
\label{eq:PaoFractionalPDE_BVP_halfspace}
\left\{
\begin{alignedat}{3}
	\big(\Id-\div(\ten{\Theta}(z)\grad)\big)^{11/12}
	\big(-\div(\ten{\Theta}(z)\grad)\big)^{1/2} \bpsi
	&=
	\mu L(z)^{17/6} \bxi_\beta
	\quad
	&&\text{in } \R_+^3,
	\\
	\kappa\sspace\psi_3 + L_3(z)^2 \frac{\partial\psi_3}{\partial z}
	=
	\psi_1 &= \psi_2 =
	0
	\quad
	&&\text{at } z=0
	.
\end{alignedat}
\right.
\end{equation}

We do not analyze these equations in detail here, however, we present a single realization of their solution~\cref{fig:VelocityField_comparison} for visual comparison.
Observe that the velocity field coming from~\cref{eq:PaoFractionalPDE_BVP_halfspace} is visibly smoother than its counterpart coming from~\cref{eq:VKFractionalPDE_BVP_halfspace}.
This is due to the high regularity load $\bxi_\beta$.

\subsection{Uniform shear boundary layers} %
\label{sub:shear_boundary_layers}

Classically, rapid distortion theory is used to describe the short time evolution of isotropic turbulence.
As pointed out in, e.g., \citet{Lee1991}, it is also possible to extend its use to some examples of inhomogeneous turbulence.
In this example, we follow \citet{Lee1991} in considering a uniform shear boundary layer (USBL) model where the only effect of the wall is to block velocity fluctuations in the normal direction.
Our derivation begins from the assumption $\langle \bfU(\bmx)\rangle = (U_0 + S x_3) \bme_1$ taken in~\cref{sub:rapid_distortion}, but we also allow for a $z$-dependent inhomogeneous diffusion tensor,
\begin{equation*}
\label{eq:DiffusionTensorRD}
	\ten{\Theta}_\tau(z) = 
	\begin{bmatrix}
	L_1(z)^2+\tau^2 L_3(z)^2 & 0 & \tau L_3(z)^2\\
	0 & L_2(z)^2 & 0 \\
	\tau L_3(z)^2 & 0 & L_3(z)^2
	\end{bmatrix}
	.
\end{equation*}
With this expression in hand, we may consider the following inhomogeneous version of~\cref{eq:general_SPDE} with $\tau = 1.0$:
\begin{equation}
\label{eq:ShearFractionalPDE_BVP}
\left\{
\begin{alignedat}{3}
	\big(\Id-\nabla\bcdot(\ten{\Theta}_\tau(z)\nabla)\big)^{17/12}\pot
	&=
	\mu L(z)^{17/6}\bmeta_\tau
	\quad
	&&\text{in } \R_+^3,
	\\
	\kappa\sspace\psi_3 + L_3(z)^2\bigg(\frac{\partial\psi_3}{\partial z} + \tau\frac{\partial\psi_3}{\partial x}\bigg)
	&=
	\psi_1 = \psi_2 =
	0
	\quad
	&&\text{at } z=0
	.
\end{alignedat}
\right.
\end{equation}

It is possible that the inhomogeneous length scales in this tensor, $L_i(z)$, may be tuned to compensate for the presence of small non-zero Reynolds stress gradients, however, we do not seek to verify that hypothesis here.
Instead, we settle for a visual comparison between the solutions of the various models.

\begin{figure}
\centering
	\includegraphics[clip=true, trim = 21cm 1cm 22cm 6cm, width=0.28\textwidth]{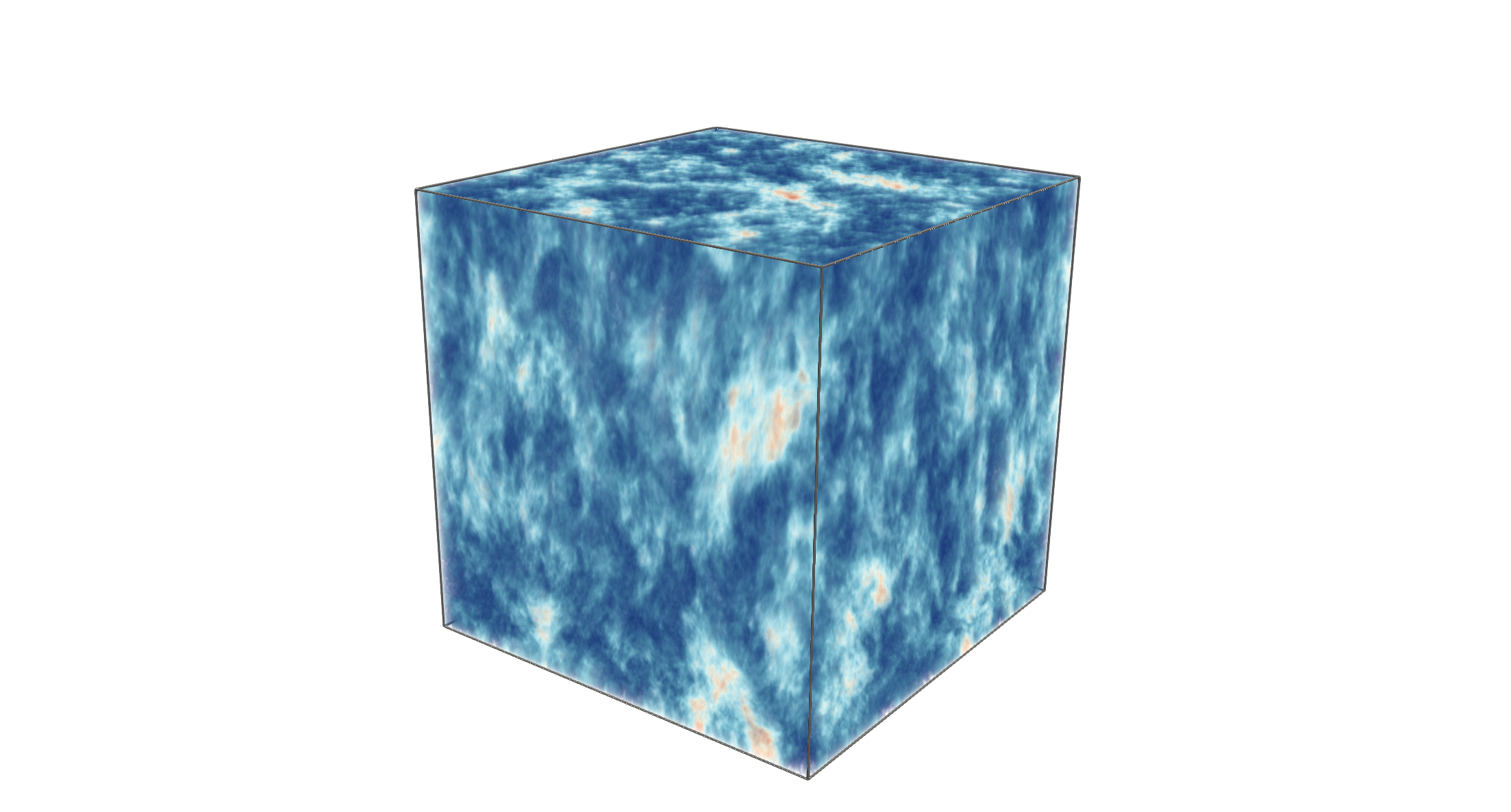}
	~
	\includegraphics[clip=true, trim = 21cm 1cm 22cm 6cm, width=0.28\textwidth]{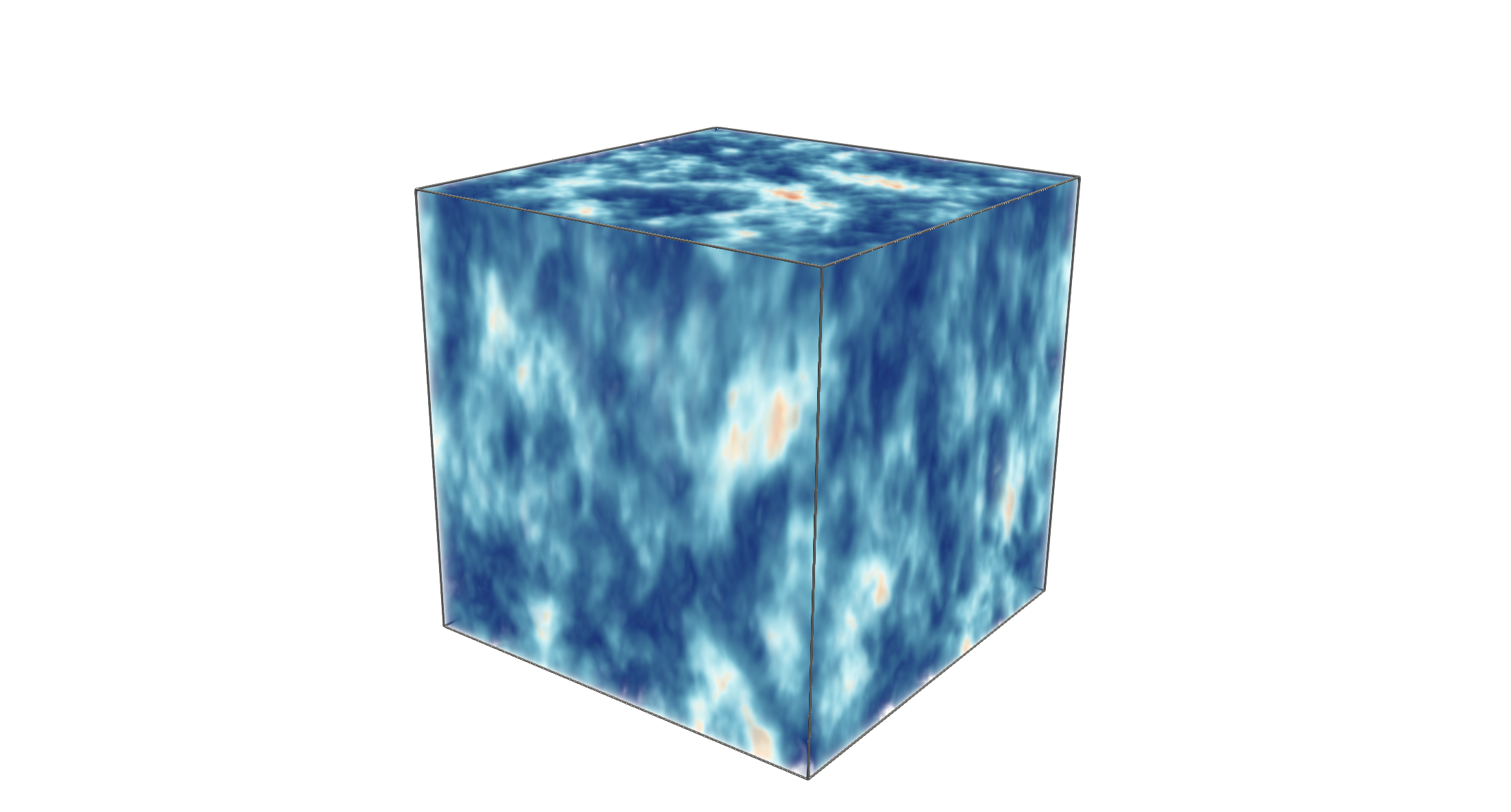}
	~
	\includegraphics[clip=true, trim = 21cm 1cm 22cm 6cm, width=0.28\textwidth]{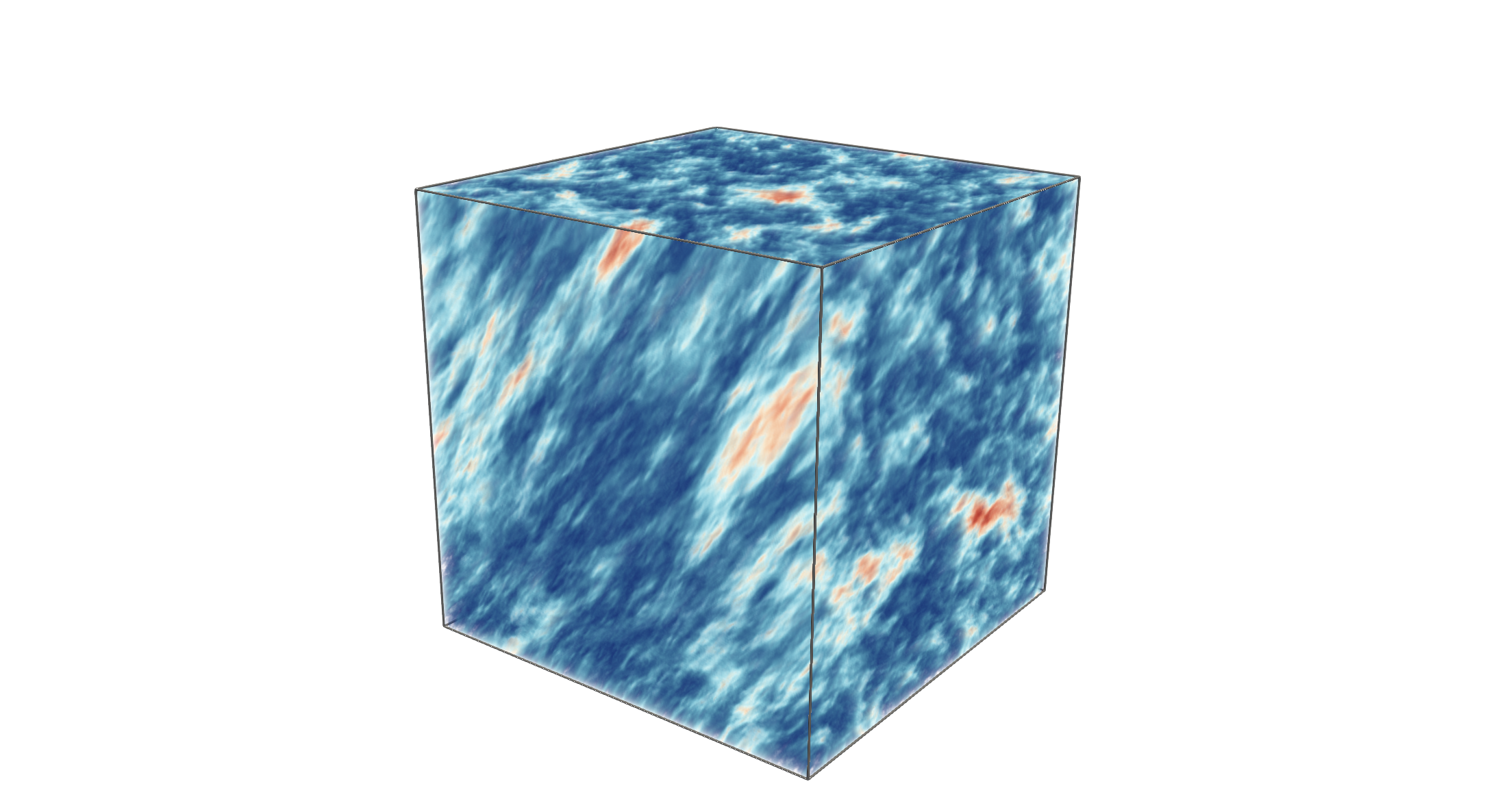}
	\caption{\label{fig:VelocityField_comparison} Magnitudes of $\bfu = \curl\bpsi$ from~\cref{eq:VKFractionalPDE_BVP_halfspace} (left),~\cref{eq:PaoFractionalPDE_BVP_halfspace} with $\beta/L_\infty = 10^{-2}$ (center), and~\cref{eq:ShearFractionalPDE_BVP} with $\tau = 1.0$ (right).
	The additional model parameters are specified in~\cref{fig:L(z)}.
	Observe that the central field is visibly smoother than its counterpart on the left due to the high regularity load $\bxi_\beta$.
	The field on the right, issued from the same noise, presents distortion.
	}
\end{figure}

\Cref{fig:VelocityField_comparison} depicts a reference velocity field coming from a single realization of~\cref{eq:VKFractionalPDE_BVP_halfspace,eq:PaoFractionalPDE_BVP_halfspace,eq:ShearFractionalPDE_BVP}.
In order to demonstrate the flexibility of the models, we have taken the same calibrated model parameters used in \cref{ssub:the_von_karman_spectrum}. 
For a fair reference, we have also used the same additive white Gaussian noise vector to generate the load for each realization.

\subsection{Turbulent inlet generation for numerical wind tunnel simulations} %
\label{sub:turbulent_inlet_generation_for_the_atmospheric_boundary_layer}

The mean profile $\langle \bfU(z) \rangle$ in many wall-bounded shear flows is often assumed to follow a logarithmic curve, sometimes with a Reynolds number modification; see, e.g., \citet{Barenblatt2004}.
In the atmospheric boundary layer, one such model for the mean velocity , $\langle\bfU(\bmx)\rangle = U(z)\bme_1$, found in the wind engineering community is written in terms of the height above ground, $z$, as follows \citep{mendis2007wind,kareem2013advanced}:
\begin{equation}
\label{eq:LogMeanProfile}
	U(z)
	=
	\frac{u_\ast}{\kappa}
	\ln\bigg(\frac{z-d}{z_0}\bigg)
	.
\end{equation}
Here, $u_\ast$ is the friction velocity, $z_0$ is the roughness length, and $d$ is the zero-plane displacement.
Although all such models violate the uniform shear assumption made in deriving~\cref{eq:RapidDistortionHomogeneousFPDE}, it has been argued that the assumption is still valid for describing eddies of ``{\it linear dimension smaller than the length over which the shear changes appreciably}'' \citep[p.~145]{mann1994spatial}.
For this reason, turbulence models similar to those presented in the previous subsections \citep[see, e.g.,][]{mann1994spatial,mann1998wind,Chougule2018}, have established themselves in wind engineering \citep{tc882005iec}.
An account of some physical violations of such models is given in detail in \citet{Hunt1984turbulence,Hunt1989Cross}.
It remains to be demonstrated whether the nonhomogenous diffusion coefficient in, e.g.,~\cref{eq:ShearFractionalPDE_BVP} may ameliorate some of these issues.

\begin{figure}
\centering
	\includegraphics[width=0.95\textwidth]{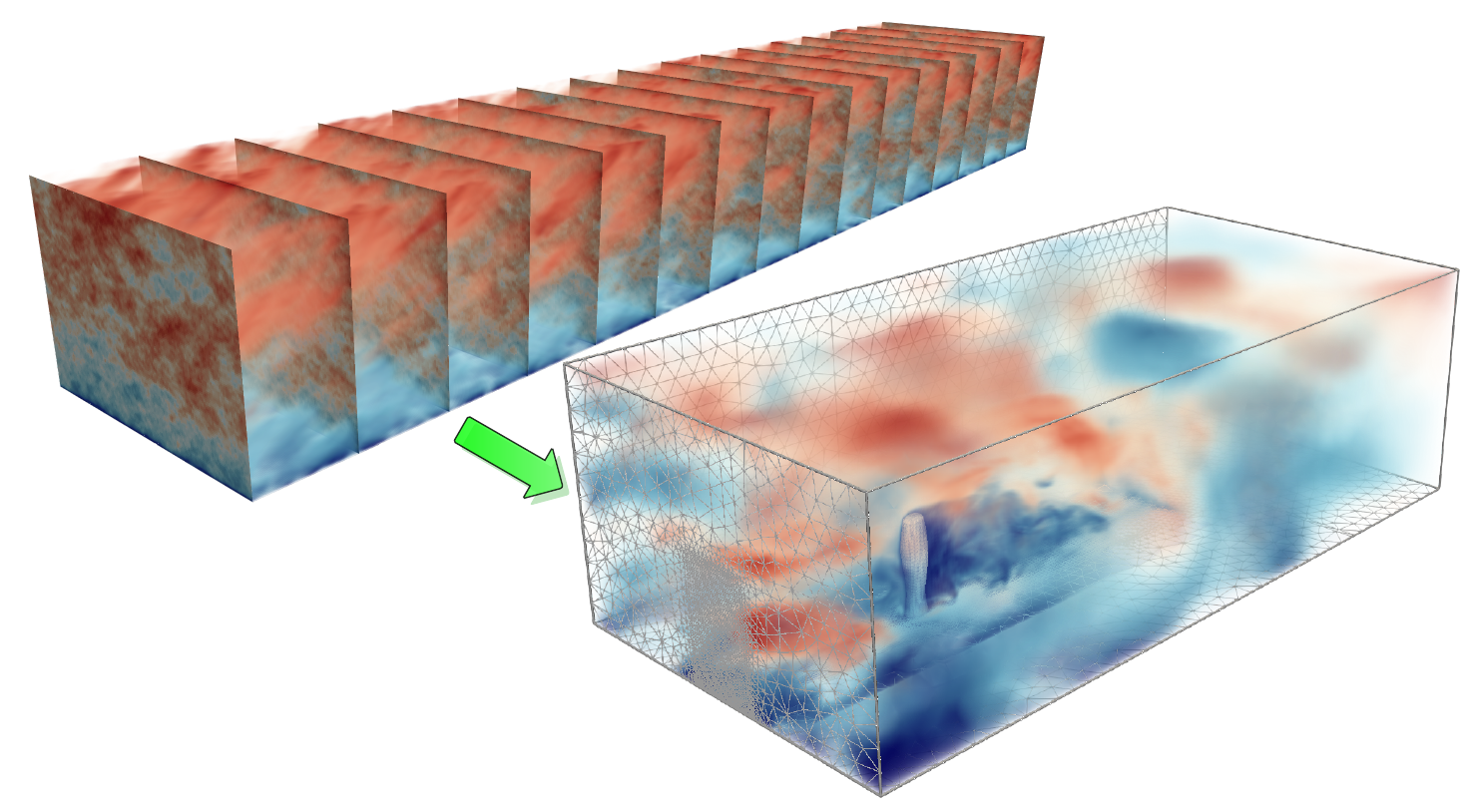}
	\caption{\label{fig:Snapshots}Snapshots of synthetic wind, $\bmU(\bmx) = \langle\bmU(\bmx)\rangle + \bfu(\bmx)$, mapped onto the inlet boundary in a numerical wind tunnel test of a modern high rise building.
	Turbulent fluctuations $\bfu(\bmx)$ generated using model~\cref{eq:ShearFractionalPDE_BVP} with $\tau = 1.0$, $\kappa = 0$, and $L_1(z) = L_2(z) = L_3(z) = L_\infty$.
	The large eddy simulation was performed with the finite element software Kratos Multiphysics \citep{dadvand2010object}.}
\end{figure}

Our final application involves using~\cref{eq:ShearFractionalPDE_BVP} to generate synthetic turbulent inlet conditions, which is an important application in CFD as a whole \citep{tabor2010inlet}.
We choose to follow an established approach used in the wind engineering industry; see \citet{Michalski2011,Andre2015} and references therein.
Here, a contiguous section of spatially correlated turbulence is transformed into a stationary Gaussian process by identifying the $x$-component of the turbulent velocity field with a time axis via the transformation $x = U_\mathrm{m} t$.
Then, at each time step $t= t_k$, the turbulent fluctuations $\bfU(\bmx)|_{x = t_k/U_\mathrm{m}}$ are projected onto the inflow boundary of a numerical wind tunnel; see depiction in~\cref{fig:Snapshots}.
Here, $U_\mathrm{m} > 0$ is a mean velocity parameter which directly affects the spatial-to-temporal correlation of the synthetic turbulent inlet boundary conditions.
With this application, we highlight the potential of calibrated FPDE models to improve the accuracy of numerical wind tunnel simulations.

\begin{remark}
	The physical justification for the transformation $x = U_\mathrm{m} t$ derives from a manipulated Taylor's hypothesis, as described in \citet[Section~2.3]{mann1994spatial}.
\end{remark}

\section{Solution and calibration} %
\label{sec:calibration_and_rational_approximation}

In this section, we briefly summarize numerical strategies for solution of fractional PDEs and, in particular, the rational approximation method which we used to solve the problems given in~\cref{sec:applications}.
We then describe how to calibrate such models so that its solutions best represent experimental data.

\subsection{Solution of fractional PDEs} %
\label{sub:rational_approximation}

Numerical solution of boundary value problems involving fractional powers of elliptic operators is challenging and computationally expensive, due in part to the non-locality of the resulting operator.
Methods based on diagonalization of the elliptic operator~\citep{ilic2005numerical,yang2011novel} are generally too expensive for practical applications.
Alternative techniques usually involve either reducing the fractional problem to a transient pseudo-parabolic problem~\citep{vabishchevich2015numerically,lazarov2017numerical} or to local elliptic problems.
The latter category includes extensions to a higher-dimensional integer-order boundary value problem on a semi-infinite cylinder~\citep{caffarelli2007extension,nochetto2015pde}, quadrature for the integral representation of the inverse operator~\citep{balakrishnan1960fractional,bonito2015numerical}, or the rational approximation of the operator's spectrum~\citep{harizanov2018positive,bolin2019rational}.
The interested reader is referred to~\citet{bonito2018numerical,lischke2020fractional} for further information on fractional diffusion problems.
In this work, we follow the rational approximation approach mentioned above.
The main idea is briefly summarized below.

Let $A$ be an abstract bounded elliptic symmetric positive definite operator with spectrum~$\sigma(A)\subset[\lambda_\mathrm{min}, \lambda_\mathrm{max}]$, $0<\lambda_\mathrm{min}< \lambda_\mathrm{max}$.
For illustration, consider the associated fractional problem
\begin{equation*}%
	A^\alpha\vec{\psi} = \vec{b},
\end{equation*}
for some $\alpha>0$.
If the rational function~$r_N(\lambda) = \sum_{n=1}^{N}\frac{c_n}{\lambda + d_n}$ approximates the function~$f(\lambda)=\lambda^{-\alpha}$ on the interval~$[\lambda_\mathrm{min}, \lambda_\mathrm{max}]$,
then the solution~$\vec{\psi}$ can be approximated as the weighted average of solutions of $N$ other elliptic problems; namely,
\begin{equation}\label{eq:RA_approximant}
\bpsi \approx
\sum_{n=1}^{N} c_n\bpsi_n,
\qquad
\big(d_n\Id + A\big)\bpsi_n
=
\bmb
.
\end{equation}
If $A$ is an integer-order differential operator, e.g., $A = \Id-\nabla\bcdot(\Theta(\bmx)\nabla)$, then each of these $N$ problems can be solved using standard discretization methods for integer-order operators, e.g., finite elements.
\cref{rem:discretization} contains a number of general comments about such discretizations.
For the reader's interest, an example of the numerical method we used for the problems in~\cref{sec:applications} is described in brief in~\cref{sec:numerical_method_for_the_half_space_domain}.

The rational approximation technique above can be extended to the solution of equation~\cref{eq:BVP} which, notably, has two fractional powers, $\alpha_1$ and $\alpha_2$.
Indeed, in this case, we need to construct a rational approximation~$r_N(\lambda)$ for the function~$f(\lambda) = \lambda^{-\alpha_1}(\lambda-1)^{-\alpha_2}$.
With this alternative rational approximation in hand, the approximate vector potential~$\tilde{\vec{\psi}}$ is again given by~\cref{eq:RA_approximant}.

\begin{remark}
\label{rem:discretization}
	Note that the load $\bmb = \mu\det(\Theta(\bmx))^{\gamma/3}\bmeta$ in~\cref{eq:BVP} is a random variable.
	The reader is referred to~\citet{lindgren2011explicit,du2002numerical,croci2018efficient} for details of numerical solution to stochastic PDEs and approximation of additive white Gaussian noise.
	Typically, a discretization of the integer-order operator equation $\big(d_n\Id + A\big)\bpsi_n = \bmb$ results in a linear system
	\begin{equation}
	\label{eq:DiscreteSystem}
	(d_n\bsfM + \bsfA)\bsfp_n = \bsfb,
	\quad
	\text{with}
	\quad
	\bsfb\sim \mcN(0,\bsfB),
	\end{equation}
	where the vector $\bsfp_n$ denotes the coefficients of the discrete solution $\bpsi^h_n$ in a preselected basis, say $\Phi$.
	Here, $\bsfM$ is a discretization of the identity operator $\Id$, $\bsfA$ is a discretization of the integer order differential operator $A$, and $\bsfB = \langle \bsfb\bsfb^\top\rangle$ is a given covariance matrix.
	Via a change of variables, the random load may also be written $\bsfb = \bsfH\bxi$, where $\bsfH\bsfH^\top = \bsfB$ and $\bxi \sim \mcN(0, \sfI)$ is a standard Gaussian vector $\langle \bxi\bxi^\top\rangle = \sfI$, with $\sfI$ denoting the identity matrix.
	One particular form of $\bsfH$ comes from the Cholesky decomposition, although many other are factorizations are also possible \citep{kessy2018optimal,croci2018efficient}.
	Finally, note that if the same basis $\Phi$ is used the solve for each $\bpsi_n^h$, then the discrete solution $\bpsi^h = \sum_{n=1}^{N} c_n\bpsi_n^h \approx \bpsi$ can also be expressed using $\Phi$, with the coefficient vector $\bsfp = \sum_{n=1}^N c_n \bsfp_n$.
\end{remark}

\begin{remark}
	The weights~$c_n$ and the poles~$-d_n$ of the rational function~$r_N(\lambda)$ can be obtained with one of the various rational approximation algorithms; see, e.g., \citet{harizanov2018positive,bolin2019rational,nakatsukasa2018aaa}.
	In this work, we used the adaptive Antoulas--Anderson (AAA) algorithm proposed in~\citet{nakatsukasa2018aaa} because of the speed and robustness we found from it in our experiments.
\end{remark}

\subsection{Fitting Reynolds stress data} %
\label{sub:fitting_data}

Various statistical quantities of a turbulent flow field can be measured experimentally.
Near a solid boundary, some of the most important of these quantities are the Reynolds stresses $\tau_{ij} = \langle u_{i}u_{j} \rangle$.
In order to calibrate the parameters in~\cref{eq:VKFractionalPDE_BVP_halfspace} to Reynolds stress data $\tau_{ij}^\text{data}(\bmx_l)$, collected at a number of locations in the flow domain $\bmx_l\in S$, we propose the following optimization problem:
\begin{equation}
\label{eq:CalibrationProblem}
	\min_\btheta
	~
	\mcJ(\btheta)
	\,,
	\quad
	\text{where}
	\quad
	\mcJ(\btheta)
	= 
	\sum_{\bmx_l\in S}
	\sum_{i,j=1}^3
	\Big(
	\tau_{ij}(\bmx_l;\btheta) - \tau_{ij}^\text{data} (\bmx_l)
	\Big)^2
	.
\end{equation}
Here, the design variable $\btheta$ denotes a coefficient vector taking accounting for all of the undetermined model parameters present in~\cref{eq:BVP}.
For instance, in~\cref{ssub:the_von_karman_spectrum} we used
\begin{equation*}
	\btheta = (c_{1,1}, d_{1,1}, c_{3,1}, d_{3,1},\ldots,c_{1,K}, d_{1,K}, c_{3,K}, d_{3,K},\kappa)
	\in \R^{4\sspace K+1}
	,
\end{equation*}
where $c_{i,k}$ and $d_{i,k}$, $i=1,3$, $k=1,\ldots,K$, appear in the representation of each $L_i(z)$ with ${K=2}$~terms; cf.~\cref{eq:Expansion}.

\begin{remark}
In turns out that~\cref{eq:CalibrationProblem} can be rewritten as a deterministic optimization problem.
To see this, recall~\cref{rem:discretization} and consider the common basis $\Phi = \{\phi_m\bme_i \,\colon m = 1,\ldots, M,~ i = 1,2,3\} \subset [H^1(\Omega)]^3$ for the discretization~\cref{eq:DiscreteSystem} of each sub-problem~\cref{eq:RA_approximant}.
We may then write $\bsfp = (\sfp_1, \ldots, \sfp_{3M})\in \R^{3M}$ and $\bpsi^h = \sum_{i=1}^3\sum_{m=1}^M \sfp_{m + (i-1)\cdot M}\phi_{m}\bme_i$.
Likewise, we may also write $\bfu^h = \sum_{i=1}^3\sum_{m=1}^M \sfp_{m + (i-1)\cdot M}\curl(\phi_{m}\bme_i)$.
As remarked previously, $\bsfp = \sum_{n=1}^{N}(d_n\bsfM + \bsfA)^{-1}c_n\bsfb$, where $\bsfb \sim \mcN(0,\bsfB)$.
Notice that both the matrices $\bsfA$ and $\bsfB$ generally depend on $\btheta$.
Throughout the rest of this section, we will use the shorthand $\bsfL^{-1}$ to denote the linear operator $\sum_{n=1}^{N}(d_n\bsfM + \bsfA)^{-1}c_n$.
With this notation at our disposal, we may simply write $\bsfp = \bsfL^{-1}\bsfb$ or, equivalently, $\bsfL\bsfp = \bsfb$.
An associated adjoint problem can be used to approximate $\tau_{ij}$ at any location $\bmx_l$.

Suppose that we wish to evaluate the covariance tensor $\langle u_i(\bmx) u_j(\bmy)\rangle$ at a point, say $\bmx_l$.
This may be approximated by applying the delta function (or some approximation thereof) in both $\bmx$- and $\bmy$-coordinates to $\langle u_i^h(\bmx) u_{j}^h(\bmy)\rangle$:
\begin{align*}
	\langle u_i^h(\bmx_l) u_{j}^h(\bmx_l)\rangle
	=
	\int_\Omega \int_\Omega \delta(\bmx-\bmx_l)\sspace \langle u_i^h(\bmx) u_{j}^h(\bmy)\rangle\sspace \delta(\bmy-\bmx_l) \dd\bmx\dd\bmy
	.
\end{align*}
Upon substitution of the expression $u_j^h = \sum_{i=1}^3\sum_{m=1}^M \sfp_{m + (i-1)\cdot M}\curl(\phi_{m}\bme_i)\bcdot \bme_j$, we find that
\begin{align*}
	\langle u_i^h(\bmx_l) u_{j}^h(\bmx_l)\rangle
	=
	\bsfd_{i,l}^\top
	\langle\bsfp\bsfp^\top\rangle
	\bsfd_{j,l}
	=
	\bsfd_{i,l}^\top \bsfL^{-1}
	\langle\bsfb\bsfb^\top\rangle
	\bsfL^{-1}
	\bsfd_{j,l}
	=
	\bsfd_{i,l}^\top \bsfL^{-1}
	\bsfB
	\bsfL^{-1}
	\bsfd_{j,l}
	,
\end{align*}
where each vector $\bsfd_{j,l} = (\sfd_{j,l,1},\ldots, \sfd_{j,l,3M})\in\R^{3M}$ is defined component-wise as $\sfd_{j,l,m+(i-1)\cdot M} = \int_\Omega \delta(\bmx-\bmx_l)\sspace\bme_j\bcdot \curl (\phi_m(\bmx)\bme_i)$ for $m=1,\ldots, M$ and $i=1,2,3$.
Hence, upon discretization, we may rewrite
\begin{equation}
\label{eq:DeterministicLoss}
	\mcJ(\btheta)
	=
	\sum_{\bmx_l\in S}
	\sum_{i,j=1}^3
	\Big(
	\bsff_{i,l}^\top\bsfB\sspace\bsff_{j,l} - \tau_{ij}^\text{data} (\bmx_l)
	\Big)^2
	,
	\quad
	\text{where each}
	\quad
	\bsfL\sspace\bsff_{i,l}
	=
	\bsfd_{i,l}
	\sspace.
\end{equation}
Because expression~\cref{eq:DeterministicLoss} is deterministic,~\cref{eq:CalibrationProblem} can be solved accurately and efficiently using a very wide variety of standard optimization software.
\end{remark}

\begin{remark}
	Owing to the fact that the loss function $\mcJ(\btheta)$ may simply be written 
	\begin{equation*}
		\mcJ(\btheta)
		=	
		\sum_{\bmx_l\in S}
		\sum_{i,j=1}^3
		\left(\bbE\left[
		u_{i}(\btheta)u_{j}(\btheta)|_{\bmx_l} - \tau_{ij}^\text{data} (\bmx_l)
		\right]\right)^2
		,
	\end{equation*}
	the optimization problem~\cref{eq:CalibrationProblem} can be solved with many stochastic optimization techniques commonly used in, e.g., the machine learning community.
	However, it is much more efficient to proceed by rewriting~\cref{eq:CalibrationProblem} as the deterministic optimization problem~\cref{eq:DeterministicLoss}.
		
	Alternatively, the optimization problem can be posed in the abstract setting of Bayesian inference. In this framework, the parameters are defined as random distributions~\citep{stuart2010inverse}.
\end{remark}

 %
\section{Conclusion} %
\label{sec:conclusions}

In this article, a class of fractional partial differential equations are presented which describe various scenarios of fully-developed wall-bounded turbulence.
Each model in this class derives from a simple ansatz on the spectral velocity tensor which, in turn, describes a wide variety of experimental data.
The various models differ from each other in the shape of their spectra in the energy-containing and dissipative ranges, in their boundary conditions (and, thus, some of their near-wall effects), in the regularity and spatial correlation of their stochastic forcing terms, and in the possible form of their diffusion tensor.

Three related applications of these models are considered.
First, calibration is performed in a shear-free boundary layer (SFBL) setting using experimental data obtained from \citet{Thomas1977}.
Here, a close match with the experimental data is clearly observed, as well as the well-known $z^{2/3}$ growth of the Reynolds stress $\langle w^2\rangle$ under a wide variety of boundary conditions.
The same calibrated model is then applied to render a turbulent velocity field in a uniform shear boundary layer (USBL).
Finally, the model is used to generate a synthetic turbulent inlet boundary condition that has inhomogeneous fluctuations in the height above ground.

The presented class of turbulence models is also compared to classical theory.
This comparison demonstrates that the FDPE description goes beyond previous methods; delivering a flexible tool for the design of new covariance models, in various flow settings, which fit experimental data.

\appendix
\section{Proofs} %
\label{app:proofs}

In this appendix, we prove~\Cref{lem:Exact_soln_ww,lem:Exact_soln_uu_N,lem:Exact_soln_uu_D}.

\begin{proof}[Proof of~\Cref{lem:Exact_soln_ww}]~
	The third velocity component is defined by $w =  \frac{\partial\psi_1}{\partial y} - \frac{\partial\psi_2}{\partial x}$, where
	\begin{gather}
		\big(\Id-L_\infty^2\Delta\big)^{\alpha} \psi_i
		=
		\mu\sspace L_\infty^{2\alpha}\sspace \xi_i
		,
		\qquad
		\psi_i\big|_{z=0} =	0
		,
		\qquad
		i=1,\,2,
		\label{eq:lemma:SPDE}
	\end{gather}
	with $\alpha = 17/12$ and $\mu=C^{1/2} \varepsilon^{1/3}$.
	Note that solutions of~\cref{eq:lemma:SPDE} can be written
	\begin{equation*}
		\psi_i(\bmx) = \int_{\R^3}\frac{\mu\,\hat{\xi}_i(\bmk)}{(1/L_\infty^2 + k_1^2 + k_2^2 + k_3^2)^{\alpha}}\, \ee^{\ii(k_1 x_1 + k_2 x_2)}\sin(k_3 x_3)\dd\bmk.
	\end{equation*}
	Hence, the third velocity component is
	\begin{equation*}
	w(\bmx) =  \mu\,\int_{\R^3}\frac{\ii k_2\hat{\xi}_1(\bmk) - \ii k_1\hat{\xi}_2(\bmk)}{(1/L_\infty^2 + |\bmk|^2)^{\alpha}}\, \ee^{\ii(k_1 x_1 + k_2 x_2)}\sin(k_3 x_3)\dd\bmk
	\end{equation*}	
	and the corresponding Reynolds stress is
	\begin{equation*}
	\avg{w^2}
	= 
	\mu^2\,\int_{\R^3}\frac{k_2^2 + k_1^2}{(1/L_\infty^2 + |\bmk|^2)^{2\alpha}}\, \sin^2(k_3 z)\dd\bmk
	,
	\end{equation*}
	since $\avg{\xi_1^2} = \avg{\xi_2^2} = 1$ and $\avg{\xi_1\xi_2} = 0$.
	Now, observe that, for any $x$, $a$, and $b$, it holds that
	\begin{equation}\label{eq:appx:proof1}
		\partial_x^2\left[\frac{1}{(a^2 + x^2)^{b-2}}\right] = \frac{4(b-2)(b-1)x^2}{(a^2 + x^2)^{b}} - \frac{2(b-2)}{(a^2 + x^2)^{b-1}}
		\,.
	\end{equation}
	Moreover, for any spatial dimension $d\geq 1$, the Fourier transform of the Mat\'ern kernel can be written~\citep[see, e.g.,][]{roininen2014whittle,Khristenko2019}
	\begin{equation}\label{eq:appx:Matern_Fourier}
		\Matern_{\nu}\left(a|\bmx|\right)
		=
		a^{2\nu}\frac{\Gamma(\nu+d/2)}{\pi^{d/2}\Gamma(\nu)}
		\int_{\R^d}\frac{1}{(a^2 + |\bmk|^2)^{\nu+d/2}} \prod_{i=1}^d\cos(x_i k_i)\dd\bmk
		.
	\end{equation}
	Therefore,
	\begin{align*}
	\avg{w^2}
	&=
	\mu^2 L_\infty^{2\nu}\,\int_{\R^3}\left[\left(\partial_{k_1}^2+\partial_{k_2}^2\right)\frac{(4(2\alpha-2)(2\alpha-1))^{-1}}{(1 + |\bmk|^2)^{2\alpha-2}} 
	+ \frac{(2\alpha-1)^{-1}}{(1 + |\bmk|^2)^{2\alpha-1}}
	\right]
	\frac{1 - \cos\left(\frac{2k_3 z}{L_\infty}\right)}{2}\dd\bmk
	\\
	&=
	\frac{\mu^2 L_\infty^{2\nu}}{2(2\alpha-1)}\,\int_{\R^3}\frac{1 - \cos\left(\frac{2k_3 z}{L_\infty}\right)}{(1 + |\bmk|^2)^{2\alpha-1}}\dd\bmk
	=
	\frac{\mu^2 L_\infty^{2\nu}}{2(\nu+d/2)}\,\frac{\pi^{d/2}\Gamma(\nu)}{\Gamma(\nu+d/2)}\,\Matern_{\nu}\left(\abs{\bmx}\right) \biggr|_{(0,0,2z/L_\infty)}^{(0,0,0)}
	\\
	&
	=
	\underbrace{\frac{\mu^2 L_\infty^{2\nu}}{2(\nu+d/2)}\,\frac{\pi^{d/2}\Gamma(\nu)}{\Gamma(\nu+d/2)}}_{=\avg{w^2_\infty}}\,\left[1 - \Matern_{\nu}\left(\frac{2z}{L_\infty}\right)\right]
	\end{align*}
	where $d=3$ and $\nu = 2\alpha-1-d/2 = 17/6 - 1 - 3/2 = 1/3$.

	Finally, the modified Bessel function of the second kind, for $\nu\notin\Z$, is defined by the expansion
	\begin{equation*}
	\BesselK_{\nu}(x) = \frac{\Gamma(\nu)\Gamma(1-\nu)}{2} \left(\sum_{m=0}^{\infty}\frac{1}{m!\,\Gamma(m-\nu+1)}\left(\frac{x}{2}\right)^{2m-\nu} - \sum_{m=0}^{\infty}\frac{1}{m!\,\Gamma(m+\nu+1)}\left(\frac{x}{2}\right)^{2m+\nu}\right).
	\end{equation*}
	Hence, we have
	\begin{equation*}
	\Matern_{\nu}\left(\frac{2z}{L_\infty}\right) \sim 1 - \frac{\Gamma(1-\nu)}{\Gamma(1+\nu)}\left(\frac{z}{L_\infty}\right)^{2\nu}
	\quad
	\text{as } \frac{z}{L_\infty} \to 0\,.
	\end{equation*}	
	From this and \cref{eq:lemma:statement}, the statement follows.	
\end{proof}

\begin{proof}[Proof of~\Cref{lem:Exact_soln_uu_N}]~
	The first two components of the vector potential~$\bpsi$ are defined by~\eqref{eq:lemma:SPDE}, while the third component is defined by
	\begin{gather}
	\big(\Id-L_\infty^2\Delta\big)^{\alpha} \psi_3
	=
	\mu\sspace L_\infty^{2\alpha}\sspace \xi_3
	,
	\qquad
	\partial_z\psi_3\big|_{z=0} = 0
	,
	\label{eq:lemma:SPDE3}
	\end{gather}
	with $\alpha = 17/12$ and $\mu=C^{1/2} \varepsilon^{1/3}$.
	Note that solutions of~\cref{eq:lemma:SPDE3} can be written
	\begin{equation*}
	\psi_3(\bmx) = \int_{\R^3}\frac{\mu\,\hat{\xi}_3(\bmk)}{(1/L_\infty^2 + |\bmk|^2)^{\alpha}}\, \ee^{\ii(k_1 x_1 + k_2 x_2)}\cos(k_3 x_3)\dd\bmk.
	\end{equation*}
	Hence, the two first velocity components are
	\begin{align*}
	u(\bmx) =  \mu\,\int_{\R^3}\frac{ \ii k_2\hat{\xi}_3(\bmk) - k_3\hat{\xi}_2(\bmk)}{(1/L_\infty^2 + |\bmk|^2)^{\alpha}}\, \ee^{\ii(k_1 x_1 + k_2 x_2)}\cos(k_3 x_3)\dd\bmk, \\
	v(\bmx) =  \mu\,\int_{\R^3}\frac{ k_3\hat{\xi}_1(\bmk) - \ii k_1\hat{\xi}_3(\bmk)}{(1/L_\infty^2 + |\bmk|^2)^{\alpha}}\, \ee^{\ii(k_1 x_1 + k_2 x_2)}\cos(k_3 x_3)\dd\bmk,
	\end{align*}	
	and the corresponding Reynolds stresses are
	\begin{align*}
	\avg{u^2} = \avg{v^2}
	&= 
	\mu^2\,\int_{\R^3}\frac{k_3^2 + k_i^2}{(1/L_\infty^2 + |\bmk|^2)^{2\alpha}}\, \cos^2(k_3 z)\dd\bmk
	,
	\qquad
	i=1\text{ or }2,
	\end{align*}
	since $\avg{\xi_1^2} = \avg{\xi_2^2} = 1$ and $\avg{\xi_1\xi_2} = 0$.
	Taking in account~\eqref{eq:appx:proof1} and~\eqref{eq:appx:Matern_Fourier}, we obtain
	\begin{align*}
	\avg{u^2}
	&=
	\mu^2 L_\infty^{2\nu}\,\int_{\R^3}\left(\frac{1}{(1 + |\bmk|^2)^{2\alpha-1}} - \frac{1+ k_i^2}{(1 + |\bmk|^2)^{2\alpha}}\right)\,
	\frac{1 + \cos\left(\frac{2k_3 z}{L_\infty}\right)}{2}\dd\bmk
	\\
	&=
	\frac{\mu^2 L_\infty^{2\nu}}{2}\,\int_{\R^3}\left(\frac{1-(2(2\alpha-1))^{-1}}{(1 + |\bmk|^2)^{2\alpha-1}} - \frac{1}{(1 + |\bmk|^2)^{2\alpha}} 
	 \right)\, 
	 \left[1 + \cos\left(\frac{2k_3 z}{L_\infty}\right)\right]\dd\bmk
	\\	
	&=
	\frac{\mu^2 L_\infty^{2\nu}\pi^{d/2}}{2}\,\left(\frac{\Gamma(\nu)}{\Gamma(\nu+\frac{d}{2})} \frac{\nu+1}{\nu+\frac{d}{2}}
	\left[1 + \Matern_{\nu}\left(\frac{2z}{L_\infty}\right)\right] - \frac{\Gamma(\nu+1)}{\Gamma(\nu+1+\frac{d}{2})}\left[1 + \Matern_{\nu+1}\left(\frac{2z}{L_\infty}\right)\right]\right)
	\\	
	&=
	\avg{u^2_\infty}\,
	\left[1 + (\nu+1)\Matern_{\nu}\left(\frac{2z}{L_\infty}\right) - \nu\Matern_{\nu+1}\left(\frac{2z}{L_\infty}\right)\right],
	\end{align*}
	where $d=3$ and $\nu = 2\alpha-1-d/2 = 17/6 - 1 - 3/2 = 1/3$, and $\avg{u^2_\infty} = \avg{w^2_\infty}$.
\end{proof}

\begin{proof}[Proof of~\Cref{lem:Exact_soln_uu_D}]~
	The components of the vector potential~$\bpsi$ are defined by equations~\eqref{eq:lemma:SPDE} and~\eqref{eq:lemma:SPDE3} with homogeneous Dirichlet boundary condition~$\psi_3\big|_{z=0} = 0$, and thus have form
	\begin{equation*}
	\psi_i(\bmx) = \int_{\R^3}\frac{\mu\,\hat{\xi}_i(\bmk)}{(1/L_\infty^2 + |\bmk|^2)^{\alpha}}\, \ee^{\ii(k_1 x_1 + k_2 x_2)}\sin(k_3 x_3)\dd\bmk, \qquad i=1,2,3.
	\end{equation*}
	Hence, the two first velocity components are
	\begin{align*}
	u(\bmx) =  \mu\,\int_{\R^3}\frac{ \ii k_2\hat{\xi}_3(\bmk)\sin(k_3 x_3) - k_3\hat{\xi}_2(\bmk)\cos(k_3 x_3)}{(1/L_\infty^2 + |\bmk|^2)^{\alpha}}\, \ee^{\ii(k_1 x_1 + k_2 x_2)}\dd\bmk, \\
	v(\bmx) =  \mu\,\int_{\R^3}\frac{ k_3\hat{\xi}_1(\bmk)\cos(k_3 x_3) - \ii k_1\hat{\xi}_3(\bmk)\sin(k_3 x_3)}{(1/L_\infty^2 + |\bmk|^2)^{\alpha}}\, \ee^{\ii(k_1 x_1 + k_2 x_2)}\dd\bmk,
	\end{align*}	
	and the corresponding Reynolds stresses are
	\begin{align*}
	\avg{u^2} = \avg{v^2}
	&= 
	\mu^2\,\int_{\R^3}\frac{k_3^2\cos^2(k_3 z) + k_i^2\sin^2(k_3 z)}{(1/L_\infty^2 + |\bmk|^2)^{2\alpha}}\, \dd\bmk
	\\
	&=
	\mu^2\,\int_{\R^3}\frac{(k_3^2+k_i^2)\cos^2(k_3 z)}{(1/L_\infty^2 + |\bmk|^2)^{2\alpha}}\, \dd\bmk
	-
	\mu^2\,\int_{\R^3}\frac{k_i^2\cos(2 k_3 z)}{(1/L_\infty^2 + |\bmk|^2)^{2\alpha}}\, \dd\bmk
	,
	\end{align*}
	since $\avg{\xi_1^2} = \avg{\xi_2^2} = 1$ and $\avg{\xi_1\xi_2} = 0$.
	Taking in account the two previous proofs, we obtain
	\begin{align*}
	\avg{u^2}
	&=
	\avg{u^2_\infty}\,
	\left(\left[1 + (\nu+1)\Matern_{\nu}\left(\frac{2z}{L_\infty}\right) - \nu\Matern_{\nu+1}\left(\frac{2z}{L_\infty}\right)\right] - \Matern_{\nu}\left(\frac{2z}{L_\infty}\right)\right)
	\\	
	&=
	\avg{u^2_\infty}\,
	\left[1 + \nu\Matern_{\nu}\left(\frac{2z}{L_\infty}\right) - \nu\Matern_{\nu+1}\left(\frac{2z}{L_\infty}\right)\right],	
	\end{align*}
	where $d=3$ and $\nu = 2\alpha-1-d/2 = 17/6 - 1 - 3/2 = 1/3$, and $\avg{u^2_\infty} = \avg{w^2_\infty}$.
\end{proof}

\section{Numerical method for the half-space domain} %
\label{sec:numerical_method_for_the_half_space_domain}

In this appendix, we deal with the numerical approximation of the boundary values problems given in~\cref{sec:applications}.
We focus at first on~\cref{eq:ShearFractionalPDE_BVP} as a representative example, as it is the most challenging.
Like all numerical approximations of problems on unbounded domains $\Omega$, we only seek to render the solution in a prespecified bounded subdomain $\Omega_0 \subsetneq \Omega$.
In practice, this also requires us to define a larger domain for computation, say $\Omega^{\mathrm{comp.}} := [0,x_{\mathrm{max}}]\times[0,y_{\mathrm{max}}]\times[0,z_{\mathrm{max}}] \subset\Omega$, containing $\Omega_0$.
If adequate care is taken in defining it, the solution $\bfu^{\mathrm{comp.}}$ of a related problem on $\Omega^{\mathrm{comp.}}$ will be close to the true solution $\bfu$, once they are both restricted to $\Omega_0\subsetneq \Omega^\mathrm{comp.}$ \citep{Khristenko2019}; i.e., $\bfu^{\mathrm{comp.}}|_{\Omega_0} \approx \bfu|_{\Omega_0}$.

Consider the solution $\bpsi(\bmx;\tau)$ of~\cref{eq:ShearFractionalPDE_BVP}.
After applying a Fourier transform in the $x$- and $y$-directions, we arrive at the transformed vector potential
\begin{equation*}
	\hat{\bpsi}(k_1,k_2,z;\tau)
	=
	\frac{1}{(2\pi)^2}
	\int_{-\infty}^\infty
	\int_{-\infty}^\infty
	\ee^{-\ii(k_1x_1 + k_2x_2)}
	{\bpsi}(x_1,x_2,z;\tau)
	\dd x_1 \dd x_2
	\,.
\end{equation*}
For each $k_1,k_2\in\R$ and $t\geq 0$, we can then rewrite~\cref{eq:ShearFractionalPDE_BVP} as a one-dimensional boundary value problem for $\hat{\bpsi} = \hat{\bpsi}(k_1,k_2,z;\tau)$, as follows:
\begin{equation}
\label{eq:ShearFractionalPDE_BVP_Fourier}
\left\{
\begin{alignedat}{3}
	\hat{A}(k_1,k_2,z;\tau)^{\alpha}\hat{\pot}
	&=
	\bmb(k_1,k_2,z;\tau)
	\quad
	&&\text{for } z>0,
	\\
	\kappa\sspace\hat{\psi}_3 + L_3(z)^2\bigg(\frac{\partial\hat{\psi}_3}{\partial z} + \ii\! k_1\tau\hat{\psi}_3\bigg)
	&=
	\hat{\psi}_1 = \hat{\psi}_2
	= 0
	\quad
	&&\text{at } z=0
	,
\end{alignedat}
\right.
\end{equation}
where $\alpha = 17/12$, $\bmb(k_1,k_2,z;\tau) = \mu L(z)^{2\alpha} \Fourier_{z}^{-1}\big[\bsfD_\tau^{-\top}\hat{\bxi}\big](k_1,k_2,z)$, and
\begin{multline}\label{eq:ODE2_dst}
	\quad
	\hat{A}(k_1,k_2,z;\tau)
	=
	\Id + (L_1(z)^2 + \tau^2 L_3(z)^2 )k_1^2 + L_2(z)^2 k_2^2
	\\ - \ii\!\tau k_1\left(L_3(z)^2  \drv{}{z} + \drv{}{z}L_3(z)^2\right) - \drv{}{z}L_3(z)^2\drv{}{z}
	.
	\quad
\end{multline}

The continuous Fourier transforms in the $x$- and $y$-directions used in deriving~\cref{eq:ShearFractionalPDE_BVP_Fourier} can be replaced by discrete Fourier transforms on uniform grids over the intervals $[0,x_{\mathrm{max}}]$ and $[0,y_{\mathrm{max}}]$, respectively.
Likewise, the equation $\hat{A}^{\alpha}\hat{\pot} = \bmb$ can be solved in a finite interval $[0,z_{\mathrm{max}}]$, once supplementary boundary conditions are applied at the artificial boundary $z = z_{\mathrm{max}}$ in order to close the resulting system of equations.
For instance, one may apply the Dirichlet boundary condition
\begin{equation*}
	\hat{\bpsi} = \bzero
	\quad
	\text{at } z=z_{\mathrm{max}}.
\end{equation*}
In our experiments, we also experimented with zero flux boundary conditions at $z=z_{\mathrm{max}}$ and witnessed similar results near the boundary $z=0$.
In general, a wide variety of different boundary conditions may be applied at the artificial interfaces/boundaries $x = x_{\mathrm{max}}$, $y = y_{\mathrm{max}}$, and $z = z_{\mathrm{max}}$, with negligible cost to solution accuracy, so long as $x_{\mathrm{max}}$, $y_{\mathrm{max}}$, and $z_{\mathrm{max}}$ are each sufficiently large; cf. \citet{Khristenko2019}.

The numerical approximation of~\cref{eq:ShearFractionalPDE_BVP} then proceeds by applying the rational approximation algorithm presented in~\cref{sub:rational_approximation} to a discrete form of~\cref{eq:ShearFractionalPDE_BVP_Fourier}, applying an inverse discrete Fourier transform in both the $k_1$- and $k_2$-coordinates, and restricting the resulting solution to $\Omega_0 \subsetneq \Omega^{\mathrm{comp.}}$.

For example, let $V_h = \spann\{\phi_1,\ldots,\phi_M\} \subset H^1_0(0,z_{\max})$ be a suitable approximation subspace (e.g., each $\phi_i$ could be a piecewise-linear hat function) and consider the special case $\tau = 0$ and $\kappa=\infty$.
We wish to compute an approximation $\bpsi^h = (\psi^h_1,\psi^h_2,\psi^h_2)\approx \bpsi$ in $V_h$.
In this setting, the basis function expansion of $\psi_i^h = \mcF^{-1}_{x,y}\big[\sum_{n=1}^N \sum_{m=1}^M \sfp_{n,(i-1)\cdot M+m}\phi_{m} \big]$, $i=1,2,3$, is determined by the solution of the $N$ linear systems,
\begin{equation*}
	\left(
	d_n
	\begin{bmatrix}
		\bsfM & 0 & 0\\
		0 & \bsfM & 0\\
		0 & 0 & \bsfM\\
	\end{bmatrix}
	+
	\begin{bmatrix}
		\bsfA & 0 & 0\\
		0 & \bsfA & 0\\
		0 & 0 & \bsfA\\
	\end{bmatrix}
	\right)
	\bsfp_n = c_n\bsfb
	,
	\quad
	\text{with}
	\quad
	\bsfb \sim \mcN\left(0, \begin{bmatrix}
						\bsfB & 0 & 0\\
						0 & \bsfB & 0\\
						0 & 0 & \bsfB\\
					  \end{bmatrix}\right)
  ,
\end{equation*}
as in~\cref{eq:DiscreteSystem}.
Here, $[\bsfM]_{lm} = \int_{0}^{z_{\max}} \phi_l(z)\phi_m(z) \dd z$,
\begin{align*}
	[\bsfA]_{lm}
	&=
	\int_{0}^{z_{\max}}
	\big(1 + L_1(z)^2 k_1^2 + L_2(z)^2 k_2^2\big) 
	\phi_l(z)\phi_m(z)
	\dd z
	+
	\int_{0}^{z_{\max}}
	L_3(z)^2\drv{\phi_l(z)}{z}\drv{\phi_m(z)}{z}
	\dd z
	,
\end{align*}
and $[\bsfB]_{lm} = \int_{0}^{z_{\max}} \mu^2 L^{4\alpha}(z) \phi_l(z)\phi_m(z) \dd z$.
Finally, the discrete vector field $\bfu^h = \curl \bpsi^h$ can be post-processed immediately using the fact that
\begin{equation*}
\label{eq:CurlFourier}
	\mcF_{x,y}\big[\curl \bpsi^h]
	=
	\begin{bmatrix}
		0 & \frac{\partial}{\partial z} & \ii\! k_2\\
		-\frac{\partial}{\partial z} & 0 & -\ii\! k_1\\
		-\ii\! k_2 & 0 & \ii\! k_1
	\end{bmatrix}
	\hat{\bpsi}^h
	.
\end{equation*}

\begin{remark}
	When the diffusion coefficients $L_i(z)$ are constant, it is possible to apply the $z$-direction Fourier transform $\Fourier_z$ to~\cref{eq:ShearFractionalPDE_BVP_Fourier}.
	In this case, the operator $\Fourier_z\big[\hat{A}(k_1,k_2,z;\tau)^{\alpha}\big]$ can be inverted algebraically and the rational approximation algorithm can be avoided.
	This fact is useful in proving~\Cref{lem:Exact_soln_ww,lem:Exact_soln_uu_N,lem:Exact_soln_uu_D}; cf.~\cref{app:proofs}.
	We hesitate to advocate for a complete discrete Fourier transform approach to numerical solution in the constant coefficient scenario because additional care is required in order to handle the Robin boundary condition $\Fourier_z\big[ \kappa\sspace\hat{\psi}_3 + \big(\frac{\partial\hat{\psi}_3}{\partial z} + \ii\! k_1\tau\hat{\psi}_3\big)\sspace L_3^2 \big] = (\kappa + \ii(k_3 + k_1\tau) L_3^2)\Fourier_z\big[\hat{\psi}_3\big] = 0$ when $\kappa\in(0,\infty)$; see \citet{daon2016mitigating,Khristenko2019} and references therein.
\end{remark}

\begin{remark}
	Experience indicates that in order to produce an accurate velocity field $\bfu = \curl \bpsi$ with the approach above, it is necessary to include high frequencies $k_1$, and $k_2$.
	This may be due in part to the slow decay rate of the energy spectrum function~\cref{eq:GeneralSpectrumAnsatz}.
\end{remark}

 \medskip

\noindent{\bf Funding\bf{.}}
This project has received funding from the European Union's Horizon 2020 research and innovation programme under grant agreement No 800898.
This work was also partly supported by the German Research Foundation by grants WO671/11-1 and WO671/15-2.
\\

\noindent{\bf Acknowledgments\bf{.}}
The first two authors wish to thank Michael~Andre for the interesting discussions we had last year on synthetic inlet boundary conditions.
Those discussions inspired many of the first ideas which led to this article.
We would also like to thank Anoop~Kodakkal, Andreas~Apostolatos, Matthew~Keller, and Dagmawi~Bekel for helping set up the numerical wind tunnel simulation featured in~\cref{fig:Snapshots}.
\\

\noindent{\bf Declaration of interests\bf{.}}
The authors report no conflict of interest.
\\

\noindent{\bf Author contributions\bf{.}}
B.K. and U.K. contributed equally to analysing data and reaching conclusions, performing simulations, and in
writing the paper.

\phantomsection%
\bibliographystyle{jfm}
\bibliography{main}

\end{document}